\documentclass[12pt, draftclsnofoot, peerreviewca, onecolumn]{IEEEtran}
\usepackage{cite}
\usepackage{graphicx}
\usepackage[cmex10]{amsmath}
\usepackage{amsfonts}
\usepackage{amssymb}
\ifCLASSOPTIONcompsoc
\usepackage[caption=false,font=normalsize,labelfont=sf,textfont=sf]{subfig}
\else
\usepackage[caption=false,font=footnotesize]{subfig}
\fi
\usepackage{color}
\usepackage{multirow}
\usepackage{cases}
\usepackage{algorithm}
\usepackage{algpseudocode}

\begin{document}

\title{Precoder Design for Multi-antenna\\Partial Decode-and-Forward
(PDF)\\Cooperative Systems with Statistical CSIT\\and MMSE-SIC
Receivers}

\newtheorem{Thm}{Theorem}
\newtheorem{Lem}{Lemma}
\newtheorem{Cor}{Corollary}
\newtheorem{Def}{Definition}
\newtheorem{Exam}{Example}
\newtheorem{Alg}{Algorithm}
\newtheorem{Prob}{Problem}
\newtheorem{Rem}{Remark}
\newtheorem{assumption}{Assumption}
\newtheorem{problem}{Problem}

\author{Eddy~Chiu, Vincent~K.~N.~Lau, Shunqing Zhang and Bao~S.~M.~Mok
\thanks{The results in this paper were presented in part at IEEE GLOBECOM'09,
December 2009.}
\thanks{E.~Chiu, V.~K.~N.~Lau, and Bao~S.~M.~Mok are with the
Department of Electronic and Computer Engineering, Hong Kong
University of Science and Technology, Hong Kong (e-mail:
echiua@ieee.org, eeknlau@ust.hk, and eebao@ust.hk).}
\thanks{S.~Zhang was with the Department
of Electronic and Computer Engineering, Hong Kong University of
Science and Technology, Hong Kong. He is now with Huawei
Technologies, Co. Ltd., China (e-mail: sqzhang@huawei.com).}}

\markboth{To appear in IEEE Transactions on Wireless
Communications,~2012} {{Chiu et al.}: Precoder Design for
Multi-antenna PDF Systems with Statistical CSIT and MMSE-SIC
Receivers}

\maketitle

\begin{abstract}
Cooperative communication is an important technology in next
generation wireless networks. Aside from conventional
amplify-and-forward (AF) and decode-and-forward (DF) protocols, the
partial decode-and-forward (PDF) protocol is an alternative relaying
scheme that is especially promising for scenarios in which the relay
node cannot reliably decode the complete source message. However,
there are several important issues to be addressed regarding the
application of PDF protocols. In this paper, we propose a PDF
protocol and MIMO precoder designs at the source and relay nodes.
The precoder designs are adapted based on statistical channel state
information for correlated MIMO channels, and matched to practical
minimum mean-square-error successive interference cancelation
(MMSE-SIC) receivers at the relay and destination nodes. We show
that under similar system settings, the proposed MIMO precoder
design with PDF protocol and MMSE-SIC receivers achieves substantial
performance enhancement compared with conventional baselines.
\end{abstract}


\newpage 

\section{Introduction}
\label{sect:intro} Cooperative communication is an important
technology in next generation wireless networks. Exploiting the
broadcast nature of wireless transmission, cooperation among
different users can be efficiently utilized to significantly
increase reliability as well as the achievable rates. In the current
literature, various relaying protocols have been proposed taking
into consideration of the duplexing constraint \cite{Laneman04}.
Conventional relaying protocols include amplify-and-forward (AF) and
decode-and-forward (DF). For AF protocols, the relay node simply
scales and forwards the received signal to the destination node. One
disadvantage of AF protocols is the noise amplification in the
process of repeating the received signal. On the other hand, for DF
protocols, the relay node forwards a clean copy of the decoded
source message to the destination node. However, the relay node only
assists with data transmission if it can reliably decode the source
message, thus resulting in under-utilization of the available
resources and performance loss.

To overcome the limitations of conventional AF and DF protocols,
various alternative relaying protocols have been proposed in the
literature. For example, in \cite{Avestimehr07}, the authors
proposed a bursty amplify-and-forward (BAF) protocol. It is
demonstrated that in the low SNR and low outage probability regime,
the achievable $\epsilon$-outage capacity\footnote{The
$\epsilon$-outage capacity is the largest data rate such that the
outage probability is less than $\epsilon$.} of the BAF protocol is
more attractive than conventional AF and DF protocols. In
\cite{Azarian05}, the authors proposed a dynamic decode-and-forward
protocol, which is shown to achieve the optimal
diversity-multiplexing tradeoff (DMT) when the multiplexing gain is
smaller than $1/2$. In \cite{Yuksel04}, the authors proposed a
partial decode-and-forward (PDF) protocol for Gaussian broadcast
channels. It is assumed that the source node employs a 2-level
superposition coding scheme, and the relay node forwards
\emph{partial} information of the source message to the destination
node depending on how much information can be decoded. The PDF
protocol is especially effective when the relay node cannot reliably
decode the complete source message, and can improve upon the
relaying efficiency of conventional DF protocols in cooperative
relay systems. While a number of works have studied the theoretical
capacity of PDF systems, several important practical issues remain
to be addressed regarding the application of PDF protocols in
multi-antenna cooperative systems.

\begin{list}{\labelitemi}{\leftmargin=0.5em}
\item{\bf Precoder Design for Cooperative PDF Systems:} It has been
shown that MIMO precoding can effectively boost the performance of
multi-antenna cooperative systems \cite{Ding08}. Many prior works
considered precoder design at the relay node only (e.g.
\cite{Jnl:RelayPrecoderOnly, Cnf:RelayPrecoderOnly}). In addition, a
common assumption is that perfect channel state information (CSI) is
available to facilitate precoder design. However, in practice, there
may only be imperfect or statistical CSI at the transmitters (CSIT).
For instance, in \cite{ImperfectCSIRelayPrecoder:HKU}, the authors
consider relay precoder design in AF relay systems under imperfect
CSIT. As we illustrate in this paper, precoder design at both the
source and relay nodes is very important to exploit the full benefit
of cooperative communication. Furthermore, most existing works on
MIMO precoder design for cooperative systems are focused on AF and
DF protocols. It is very challenging to design the source and relay
node precoders tailoring to the PDF protocol with statistical CSIT.

\item{\bf Precoder Design Matched to Practical MMSE-SIC Receivers:}
Precoder design is tightly coupled with the receiver structure at
the relay and destination nodes. In \cite{Khoshnevis08}, the authors
consider precoder design for single-stream cooperative system with
maximal ratio combining receivers. On the other hand, traditional
precoder design for multi-stream cooperative systems assumed either
maximum likelihood (ML) receivers \cite{Lokesh08} or simple linear
minimum mean-square-error (LMMSE) receivers \cite{Sezgin07}. The ML
receiver assumption allows for simple precoder design, but ML
receivers are difficult to implement in practice and should serve as
a performance upper bound. On the other hand, the performance with
LMMSE receivers is usually too inferior to the performance with ML
receivers. It is well-known that the MMSE successive interference
cancelation (MMSE-SIC) receiver is an important practical
low-complexity receiver that could bridge the performance gap
between ML and LMMSE receivers. While most existing literature
consider the power allocation problem to maximize the SINR based
system performance \cite{Meng07}, very few works have addressed the
problem of designing precoders matched to SIC type receivers to
enhance system capacity.
\end{list}

In this paper, we propose a MIMO precoder design that is matched to
MMSE-SIC receivers for correlated multi-antenna cooperative systems
with PDF protocol. Specifically, we consider precoder designs at
both the source and relay nodes, given only statistical CSIT, to
minimize the average end-to-end per-stream outage probability. We
find that under similar system settings, the proposed MIMO precoder
design with PDF relay protocol and MMSE-SIC receivers achieves
substantial performance enhancement compared with the use of
conventional AF and DF relay protocols.

\textbf{Notations:} In the sequel, we adopt the following notations.
$\mathbb{C}_{}^{M \times N}$ denotes the set of complex $M \times N$
matrices. Upper and lower case bold letters denote matrices and
vectors, respectively. $\textrm{diag}( x_{1}^{}, \ldots, x_{N}^{} )$
denotes a diagonal matrix with the elements $x_{1}^{}, \ldots,
x_{N}^{}$ along the diagonal. $\textrm{vec}( \textbf{X} )$ denotes
the column vector obtained by stacking the columns of $\textbf{X}$.
$[\, \textbf{X}_{1}^{} \,;\, \ldots \,;\, \textbf{X}_{N}^{} \,]$
denotes the matrix obtained by vertically concatenating
$\textbf{X}_{1}^{}, \ldots, \textbf{X}_{N}^{}$.
$[\textbf{X}]_{(a:b,c:d)}^{}$ denotes the $a$-th to the $b$-th row
and the $c$-th to the $d$-th column of $\textbf{X}$. $( \cdot
)_{}^{T}$, $( \cdot )_{}^{\dag}$, and $( \cdot )_{}^{\ast}$ denote
transpose, Hermitian transpose, and conjugate, respectively. $\det (
\cdot )$,  $\textrm{Tr} \, ( \cdot )$, and $|| \cdot ||$ denote the
determinant, the trace, and the Frobenius norm of a matrix,
respectively. $\mathbf X \otimes \mathbf Y$ denotes the Kronecker
product of $\mathbf X$ and $\mathbf Y$. $\textbf{0}_{M \times N}^{}$
denotes an $M \times N$ matrix of zeros, and $\textbf{I}_{N}^{}$
denotes an $N \times N$ identity matrix. $\mathbb{E}( \cdot )$
denotes expectation. $\textrm{Pr} \, ( X | Y )$ denotes the
probability of event $X$ given event $Y$.

\section{System Model}\label{sect:chan_mod}

\subsection{Cooperative Transmission Signal Model} \label{sect:coop_trans}
\begin{figure}
\centering
\includegraphics[width = 5in]{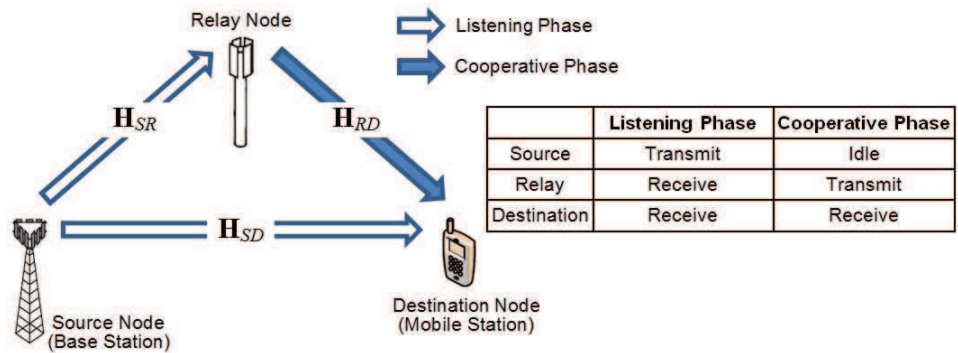}
\caption{Multi-antenna PDF cooperative system model.}
\label{fig:sys_mod}
\end{figure}

We consider a cooperative system with one source node (a base
station), one relay node, and one destination node (a mobile
station) as shown in Fig.~\ref{fig:sys_mod}. The source node is
equipped with $n_{S}^{}$ antennas, the relay node is equipped with
$n_{R}^{}$ antennas, and the destination node is equipped with
$n_{D}^{}$ antennas. The source node sends multiple independent data
streams to the destination node with the assistance of the relay
node, where both the source and relay nodes employ spatial
multiplexing transmission. The relay node operates in a half-duplex
manner, and data transmission consists of two phases: first, in the
{\em listening phase}, the source node broadcasts the data to the
relay and destination nodes; second, in the {\em cooperative phase},
the relay node forwards the source data to the destination node.
Each of the listening and cooperative phases lasts for $T$ symbol
time slots, and the end-to-end transmission lasts for a transmit
time interval (TTI) of $2T$ symbol time slots. We make the following
assumptions about the wireless channels.
\begin{assumption}[Channel Coherence Properties]\label{Assumption:FadingModel}
We assume quasi-static frequency flat fading channels, whereby the
channel coefficients of all links remain unchanged within each TTI
and varies independently from TTI to TTI.~ \hfill\IEEEQEDclosed
\end{assumption}

For each TTI, let $\mathbf s_{}^{(1)}, \ldots, \mathbf s_{}^{(Q)}$
denote the source data streams. Each data stream is separately
encoded with an inner space-time block code (STBC)
\cite{Alamouti98,OSTBC,QOSTBC} and multiplexed into the data symbols
$\mathbf X \in \mathbb C_{}^{\,n_{\!S}^{} \times T}$. Let $\mathbf
X_{R}^{} \in \mathbb C_{}^{\,n_{\!R}^{} \times T}$ denote the relay
node data symbols. The entries of $\mathbf X$ and $\mathbf X_{R}^{}$
are normalized to have unit average power. Prior to
transmission\footnote{As detailed in
Assumption~\ref{Assumption:Csi}, the transmitters only have
statistical channel knowledge. Joint STBC and precoding can be
applied to best exploit the available CSI
\cite{Jnl:RobustQosP2PMimo:Palomar,
Std:16m}.}\newcounter{StbcFootnote}\setcounter{StbcFootnote}{\value{footnote}},
the source node precodes the data symbols $\mathbf X$ using the
precoder $\mathbf P_{\!S}^{} \in \mathbb C_{}^{\,n_{\!S}^{} \times
n_{\!S}^{}}$, and the relay node precodes the data symbols $\mathbf
X_{R}^{}$ using the precoder $\mathbf P_{\!R}^{} \in \mathbb
C_{}^{\,n_{\!R}^{} \times n_{\!R}^{}}$. Let $\mathbf H_{SD}^{} \in
\mathbb C_{}^{\,n_{\!D}^{} \times n_{\!S}^{}}$ denote the channel
matrix of the source-destination (SD) link, let $\mathbf H_{SR}^{}
\in \mathbb C_{}^{\,n_{\!R}^{} \times n_{\!S}^{}}$ denote the
channel matrix of the source-relay (SR) link, and let $\mathbf
H_{RD}^{} \in \mathbb C_{}^{\,n_{\!D}^{} \times n_{\!R}^{}}$ denote
the channel matrix of the relay-destination (RD) link. Accordingly,
the received signals of the destination and relay nodes are given by
\begin{subnumcases}{\textrm{Listening
Phase: }\label{Eqn:ListeningPhaseReceivedSignals}} \mathbf
Y_{\!D,L}^{} = \mathbf H_{SD}^{} \, \mathbf P_{\!S}^{} \, \mathbf X
+ \mathbf Z_{\!D,L}^{} &for the destination
node\label{Eqn:ListeningPhaseReceivedSignalsD}\\
\mathbf Y_{\!R,L}^{} = \mathbf H_{SR}^{} \, \mathbf P_{\!S}^{} \,
\mathbf X + \mathbf Z_{\!R,L}^{} &for the relay
node\label{Eqn:ListeningPhaseReceivedSignalsR}
\end{subnumcases}\nopagebreak[4]
\begin{numcases}{\textrm{Cooperative Phase: }} \!\!\mathbf Y_{\!D,C}^{}
= \mathbf H_{RD}^{} \, \mathbf P_{\!R}^{} \, \mathbf X_{R}^{} +
\mathbf Z_{\!D,C}^{} &for the destination
node,~~~~\label{Eqn:ForwardingPhaseReceivedSignals}
\end{numcases}
where $\mathbf Z_{\!D,L}^{}, \mathbf Z_{\!D,C}^{} \in \mathbb
C_{}^{\,n_{\!D}^{} \times T}$ and $\mathbf Z_{\!R,L}^{} \in \mathbb
C_{}^{\,n_{\!R}^{} \times T}$ are AWGN with zero mean and variance
$N_{0}^{}$. We make the following assumptions about the channel
knowledge available at each node.
\begin{assumption}[Availability of Channel Knowledge]\label{Assumption:Csi}~
\hfill~
\begin{list}{\labelitemi}{\leftmargin=0.5em}
\item\emph{Statistical CSI at the Transmitters\footnote{By observing
the reverse channels, the source node can estimate the channel
statistics of the SD and SR links, and the relay node can estimate
the channel statistics of the RD link. Moreover, the source node can
acquire the channel statistics of the RD link via low-overhead
periodic feedback from the relay node.}:} The source node has
statistical knowledge of the channel matrices of all links. The
relay node has statistical knowledge of the RD link channel matrix.
\item\emph{Instantaneous CSI at the Receivers\footnote{The received
channel matrix can be accurately estimated, for example, in a
training phase by using a preamble.}:} The relay node has knowledge
of the precoded channel matrix of the SR link $\mathbf H_{SR}^{} \,
\mathbf P_{\!S}^{}$. The destination node has knowledge of the
precoded channel matrices of the SD and RD links $\{ \mathbf
H_{SD}^{} \, \mathbf P_{\!S}^{}, \mathbf H_{RD}^{} \, \mathbf
P_{\!R}^{} \}$.~ \hfill\IEEEQEDclosed
\end{list}
\end{assumption}

The source and relay nodes derive the precoders $\mathbf P_{\!S}^{}$
and $\mathbf P_{\!R}^{}$ based on statistical CSI (cf.
Section~\ref{sect:sol}). The relay and destination nodes employ
MMSE-SIC receivers, where the destination node combines the received
signals in the listening and cooperative phases to decode the source
data.

\textbf{Relay Protocol:} The form of the relay node data symbols
$\mathbf X_{R}^{}$ depends on the relay protocol adopted by the
cooperative system. Conventional relay protocols can be categorized
as AF and DF. For AF protocols, the relay node data symbols are
given by the scaled received signals,
\begin{IEEEeqnarray*}{l}
\textrm{AF: } \mathbf X_{R}^{} = \mathbf Y_{\!R,L}^{}
\mathbf{\Gamma}_{\!R,L}^{},\IEEEyesnumber\label{Eqn:AfRelaySymbol}
\end{IEEEeqnarray*}
where $\mathbf{\Gamma}_{\!R,L}^{} \in \mathbb C_{}^{\,T \times T}$
is a scaling matrix. In particular, to normalize the entries of
$\mathbf X_{R}^{}$ to have unit average power, the scaling matrix is
given by $\mathbf{\Gamma}_{\!R,L}^{} = \textrm{diag}\Big(
\frac{n_{\!R}^{}}{||\, [ \mathbf Y_{\!R,L}^{} ]_{(:,1)}^{} ||},
\ldots, \frac{n_{\!R}^{}}{||\, [ \mathbf Y_{\!R,L}^{} ]_{(:,T)}^{}
||} \Big)$. For DF protocols, the relay node attempts to decode the
source data. If the source data is correctly decoded\footnote{In
practice, cyclic redundancy check (CRC) is usually applied to
validate correct data decoding \cite[Section~16.3.11.1.1]{Std:16m}.
If the CRC check passes, the data is correctly decoded with
negligible checking errors (e.g. less than 0.1\%).}, the relay node
transmits the \emph{regenerated} data symbols $\widetilde{\mathbf X}
\in \mathbb C_{}^{\,n_{\!R}^{} \times T}$ to the destination node.
Conversely, if the source data is incorrectly decoded, the relay
node does not transmit. Therefore, the relay node data symbols are
given by
\begin{subnumcases}{\textrm{DF: }\mathbf X_{R}^{}=\label{Eqn:DfRelaySymbol}}
\widetilde{\mathbf X}\!\!\!\! &if the source data is
\emph{correctly} decoded\label{Eqn:DfRelaySymbolCorrect}\\
\textbf{0}_{n_{\!R}^{} \times T}^{}\!\!\!\! &if the source data is
\emph{incorrectly} decoded.\label{Eqn:DfRelaySymbolIncorrect}
\end{subnumcases}
In Section~\ref{Sec:PDF}, we propose a PDF protocol that addresses
the deficiencies of the conventional AF and DF protocols.

\subsection{Correlated MIMO Channel Model}
We consider correlated MIMO fading channels that reflect practical
communication systems. Specifically, we assume Rayleigh fading with
separable correlation properties on the two ends of the
link\footnote{As shown in \cite{Weichselberger06} and references
therein, this assumption can well describe general environments.},
and the channel matrices can be represented using the Kronecker
model.
\begin{Def}[Kronecker Model]\label{Def:Kronecker}
The channel matrix $\mathbf H \in \mathbb C_{}^{M \times N}$ can be
represented as
$\mathbf H = ( \mathbf \Lambda_{\,r}^{} )_{}^{\!1\!\!\:/\!\:2}
\mathbf G ( ( \mathbf \Lambda_{\,t}^{} )_{}^{\!1\!\!\:/\!\:2}
)_{}^{T}$, where $\mathbf \Lambda_{\,t}^{} \in \mathbb C_{}^{N
\times N}$ and $\mathbf \Lambda_{\,r}^{} \in \mathbb C_{}^{M \times
M}$ are the transmit- and receive-side correlation matrices, and
$\mathbf G \in \mathbb C_{}^{M \times N}$ is a random matrix whose
entries are independent and identically distributed (i.i.d.) as
complex Gaussian with zero mean and unit variance.~
\hfill\IEEEQEDclosed
\end{Def}

By the Kronecker model, the channel matrices of the different links
in the cooperative system can be represented as
\begin{IEEEeqnarray*}{ll}
\mathbf H_{SD}^{} = ( \mathbf \Lambda_{SD,r}^{}
)_{}^{\!1\!\!\:/\!\:2} \mathbf G_{\!SD}^{} ( ( \mathbf
\Lambda_{SD,t}^{} )_{}^{\!1\!\!\:/\!\:2} )_{}^{T},\; \mathbf
H_{SR}^{} = ( \mathbf \Lambda_{SR,r}^{} )_{}^{\!1\!\!\:/\!\:2}
\mathbf G_{\!SR}^{} ( ( \mathbf \Lambda_{SR,t}^{}
)_{}^{\!1\!\!\:/\!\:2}
)_{}^{T}\\
\mathbf H_{RD}^{} = ( \mathbf \Lambda_{RD,r}^{}
)_{}^{\!1\!\!\:/\!\:2} \mathbf G_{\!RD}^{} ( ( \mathbf
\Lambda_{RD,t}^{} )_{}^{\!1\!\!\:/\!\:2}
)_{}^{T},\IEEEyesnumber\label{Eqn:ChannelDecompose}
\end{IEEEeqnarray*}
where $\mathbf \Lambda_{SD,t}^{} \in \mathbb C_{}^{\,n_{\!S}^{}
\times \,n_{\!S}^{}}$ and $\mathbf \Lambda_{SD,r}^{} \in \mathbb
C_{}^{\,n_{\!D}^{} \times \,n_{\!D}^{}}$ are the transmit- and
receive-side correlation matrices of the SD link, $\mathbf
\Lambda_{SR,t}^{} \in \mathbb C_{}^{\,n_{\!S}^{} \times
\,n_{\!S}^{}}$ and $\mathbf \Lambda_{SR,r}^{} \in \mathbb
C_{}^{\,n_{\!R}^{} \times \,n_{\!R}^{}}$ are the transmit- and
receive-side correlation matrices of the SR link, and $\mathbf
\Lambda_{RD,t}^{} \in \mathbb C_{}^{\,n_{\!R}^{} \times
\,n_{\!R}^{}}$ and $\mathbf \Lambda_{RD,r}^{} \in \mathbb
C_{}^{\,n_{\!D}^{} \times \,n_{\!D}^{}}$ are the transmit- and
receive-side correlation matrices of the RD link\footnote{The
transmit and receive correlation matrices of each link depend on the
statistical antenna array features at the transmitter and the
receiver, respectively.}.

Without loss of generality, we assume that the channel matrices $\{
\mathbf H_{SD}^{}, \mathbf H_{SR}^{}, \mathbf H_{RD}^{} \}$ include
the effect of path loss. We discuss a typical operating scenario in
the following remark.

\begin{figure}
\centering
\includegraphics[width = 3.5in]{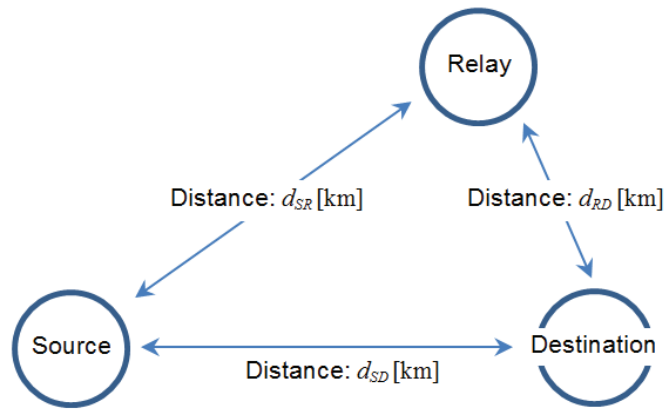}
\caption{Typical operating scenario. The path loss exponent of the
SR link is smaller than those of the SD and RD links. Moreover, the
destination node is located closer to the relay node than to the
source node. The path loss model for the SR link is given by
$-52.4-26\log_{10}^{}(d)~\textrm{[dB]}$, the path loss model for the
SD and RD links is given by $-52.4-30\log(d)\textrm{[dB]}$, where
$d$ is the distance in km between the nodes. The distance of the RD
link is smaller than the SD link, so the path loss of the RD link is
smaller than the SD link.} \label{fig:link_budget}
\end{figure}

\begin{Rem}[Typical Operating
Scenario]\label{Rem:RemarkLinkConditions} In practice, the source
node (a base station) and the relay node are mounted on rooftops,
whereas the destination node (a mobile station) is at street level.
The effect of path loss for the SD link is usually quite severe.
Since there is relatively low blockage between the source and relay
nodes, the path loss exponent of the SR link is smaller than that of
the SD link, and so the SR link is much stronger than the SD link.
Moreover, since the destination node always picks a \emph{nearby}
relay node to serve itself, the path loss of the RD link is smaller
than that of the SD link, and so the RD link is much stronger than
the SD link. We illustrate this typical operating scenario in
Fig.~\ref{fig:link_budget}.~ \hfill\IEEEQEDclosed
\end{Rem}

\section{Proposed PDF Protocol and Precoder Design Problem Formulation}
\label{sect:dsttd} A number of prior works have focused on
theoretical performance characterizations of PDF protocols (e.g.
\cite{Yuksel04}). Instead, in this paper our focus is driven by
practical considerations: we propose a PDF protocol and precoder
designs that are matched to practical receiver structures, and we
characterize the performance of such a system. In the following, we
first present the proposed PDF protocol and elaborate its advantages
over conventional AF and DF protocols. We then formulate the problem
of designing the source and relay node precoders with only
statistical CSI at the transmitters. The proposed algorithm is
suitable for scenarios in which the relay node cannot reliably
decode the complete source message (i.e. when the SR link is not
very strong). The PDF protocol and MIMO precoder design lead to a
higher probability that the relay node can assist with data
transmission and achieve non-uniform diversity protection among
spatially multiplexed streams that facilitates successive
interference cancelation decoding.

\subsection{Proposed PDF Protocol}\label{Sec:PDF}
Note that conventional relay protocols have the following key
deficiencies.
\begin{list}{\labelitemi}{\leftmargin=0.5em}
\item For conventional DF protocols, the relay node forwards the source
data only if it can correctly decode all the constituent independent
data streams. Otherwise, the relay node does not assist with data
transmission, thus compromising the achievable cooperative diversity
gain.
\item With the use of practical MMSE-SIC receivers at
the relay and destination nodes (cf. Section~\ref{sect:coop_trans}),
data decoding performance is intricately impacted by the decoding
order among the data streams \cite{Foschini99}. Specifically, the
data streams that are decoded earlier are interfered by the data
streams that are decoded subsequently. And yet, decoding errors are
propagated over successively decoded data streams. For conventional
AF and DF protocols, in the cooperative phase the relay node
forwards all (or none) of the source data streams to the destination
node. After the destination node combines the received signals in
the listening and cooperative phases, all the data streams have the
same diversity protection and experience the same order of
inter-stream interference. This makes it non-trivial to deduce the
decoding order among the data streams that would yield satisfactory
\emph{overall} data decoding performance.
\end{list}

To ameliorate the deficiencies of conventional relay protocols, we
propose a PDF protocol whereby the relay node attempts to decode and
forward a \emph{partial} of the source data streams. For notational
convenience, we introduce the following definitions.
\begin{Def}[Cooperative Streams and Regular Payload Streams]
We define the source data streams that are forwarded by the relay
node as the {\em cooperative streams}, and define the source data
streams that are not forwarded by the relay node as the {\em regular
payload streams}.~ \hfill\IEEEQEDclosed
\end{Def}

There is a higher probability that the relay node can correctly
decode a partial of the source data streams than it can correctly
decode all the data streams. As such, there is a higher probability
that the relay node can assist with data transmission based on the
proposed PDF protocol than based on conventional DF protocols.
Furthermore, after the destination node combines the received
signals in the listening and cooperative phases, the cooperative
streams that benefit from cooperative spatial diversity have higher
diversity protection compared with the regular payload streams. This
creates non-uniform diversity protection among the source data
streams; the cooperative streams can be reliably decoded first and
their interference compensated for, thus the regular payload streams
can be decoded free from interference from the cooperative streams.

\begin{figure}
\centering
\includegraphics[width = 4in]{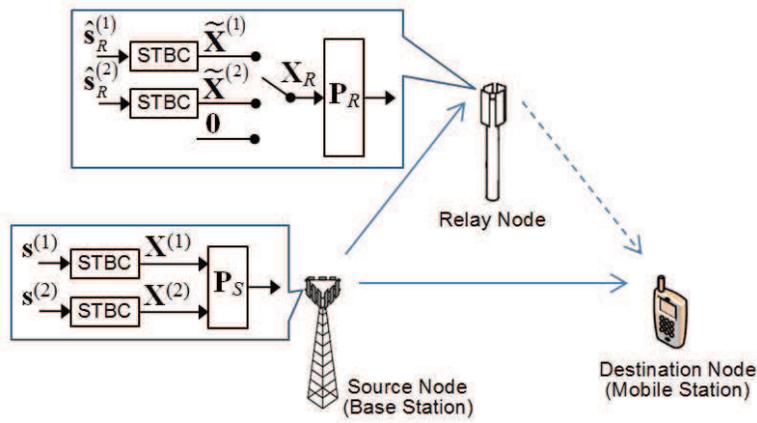}
\caption{An illustrative example of the proposed PDF protocol. The
source node sends two independent data streams to the destination
node with the assistance of the relay node. The data streams are
space-time block coded and precoded prior to transmission.}
\label{fig:trans_arch}
\end{figure}

For ease of exposition, we present the details of the proposed PDF
protocol below focusing on an illustrative scenario where the source
node sends two independent data streams to the destination node (cf.
Fig.~\ref{fig:trans_arch}).

\textbf{Processing at the Source Node:} Let $\mathbf s_{}^{(1)}$ and
$\mathbf s_{}^{(2)}$ denote the source data streams. Each data
stream is separately encoded using STBC: $\mathbf s_{}^{(1)}$ is
encoded into the symbols $\mathbf X_{}^{(1)} \in \mathbb
C_{}^{\,n_{\!S}^{(1)} \times T}$, $\mathbf s_{}^{(2)}$ is encoded
into the symbols $\mathbf X_{}^{(2)} \in \mathbb
C_{}^{\,n_{\!S}^{(2)} \times T}$ (where $n_{S}^{(1)} + n_{S}^{(2)} =
n_{S}^{}$), and $\mathbf X_{}^{(1)}$ and $\mathbf X_{}^{(2)}$ are
multiplexed into the source node data symbols $\mathbf X_{}^{} = [\;
\mathbf X_{}^{(1)} \,; \mathbf X_{}^{(2)} \;]$.

\begin{figure}
\centering
\includegraphics[width = 3.5in]{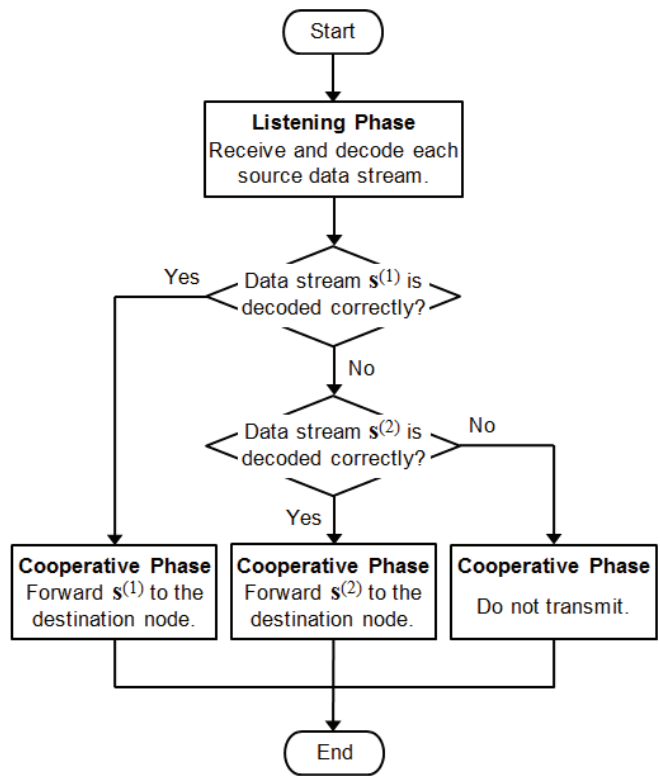}
\caption{Flow chart of the processing at the relay node. The source
node sends two independent data streams to the destination node. The
relay node attempts to decode and forward one data stream to the
destination node.} \label{fig:fc_relay}
\end{figure}

\textbf{Processing at the Relay Node:} In the listening phase, the
relay node receives the signals $\mathbf Y_{\!R,L}^{}$ and attempts
to decode one source data stream. In the cooperative phase, the
relay node forwards the correctly decoded data stream. The
processing at the relay node is illustrated in
Fig.~\ref{fig:fc_relay} and results in three cases.
\begin{list}{\labelitemi}{\leftmargin=0.5em}
\item{\em PDF~Case~1 ($\mathcal A_{1}^{}$):} In the listening phase,
the relay node correctly decodes data stream $\mathbf s_{}^{(1)}$.
In the cooperative phase, the relay node forwards $\mathbf
s_{}^{(1)}$ to the destination node.
\item{\em PDF~Case~2 ($\mathcal A_{2}^{}$):} In the listening phase,
the relay node incorrectly decodes data stream $\mathbf s_{}^{(1)}$
but correctly decodes data stream $\mathbf s_{}^{(2)}$. In the
cooperative phase, the relay node forwards $\mathbf s_{}^{(2)}$ to
the destination node.
\item {\em PDF~Case~3 ($\mathcal A_{3}^{}$):} In the listening
phase, the relay node incorrectly decodes both data streams $\mathbf
s_{}^{(1)}$ and $\mathbf s_{}^{(2)}$.  In the cooperative phase, the
relay node does not transmit.
\end{list}

The cooperative stream is encoded using STBC: when forwarding data
stream $\mathbf s_{}^{(1)}$ it is encoded into the symbols
$\widetilde{\mathbf X}_{}^{(1)} \in \mathbb C_{}^{\,n_{\!R}^{}
\times T}$, and when forwarding data stream $\mathbf s_{}^{(2)}$ it
is encoded into the symbols $\widetilde{\mathbf X}_{}^{(2)} \in
\mathbb C_{}^{\,n_{\!R}^{} \times T}$. Therefore, the relay node
data symbols are given by\footnote{In the next section, we shall
formulate the problem of designing the source node precoder $\mathbf
P_{\!S}^{}$ and the relay node precoder $\mathbf P_{\!R}^{}$ to
enhance the end-to-end outage performance of $\mathbf s_{}^{(1)}$
and $\mathbf s_{}^{(2)}$.}
\begin{subnumcases}{\textrm{PDF: }\mathbf X_{R}^{}=\label{Eqn:PdfRelaySymbol}}
\widetilde{\mathbf X}_{}^{(1)}\!\!\!\! &if $\mathbf
s_{}^{(1)}$ is \emph{correctly} decoded\\
\widetilde{\mathbf X}_{}^{(2)}\!\!\!\! &if $\mathbf s_{}^{(1)}$ is
\emph{incorrectly} decoded and $\mathbf s_{}^{(2)}$
is \emph{correctly} decoded\\
\textbf{0}_{n_{\!R}^{} \times T}^{}\!\!\!\! &if both
$\mathbf s_{}^{(1)}$ and $\mathbf s_{}^{(2)}$ are \emph{incorrectly}
decoded.
\end{subnumcases}

\textbf{Processing at the Destination Node:} The destination node
combines the received signals in the listening phase $\mathbf
Y_{\!D,L}^{}$ and the received signals in the cooperative phase
$\mathbf Y_{\!D,C}^{}$. We first decode the cooperative stream that
has higher diversity protection
, then after interference cancelation we decode the
regular payload stream
. For instance,
suppose data stream $\mathbf s_{}^{(1)}$ is the cooperative stream
(i.e. PDF~Case~1), we first decode $\mathbf s_{}^{(1)}$ and
compensate for its interference then decode data stream $\mathbf
s_{}^{(2)}$.

\subsection{Precoder Design Problem Formulation}
To enhance performance, at the source and relay nodes we employ
precoders that are designed to complement the proposed PDF protocol.
In particular, we focus on the scenario where the source node sends
two independent data streams to the destination node as previously
depicted. Suppose each of the source data streams $\mathbf
s_{}^{(1)}$ and $\mathbf s_{}^{(2)}$ contains $L$ information
bits\footnote{Given only statistical CSI at the transmitters (cf.
Assumption~\ref{Assumption:Csi}), it is not feasible to accurately
perform data rate control. We consider fixed data rate that is, for
example, determined by upper-layer protocols.}, and the data streams
are separately channel coded. With strong channel coding (such as
convolutional turbo codes (CTC) and low-density parity-check (LDPC)
codes), data decoding errors occur due to channel outage. We seek to
design the source node precoder $\mathbf P_{\!S}^{}$ and the relay
node precoder $\mathbf P_{\!R}^{}$ to minimize the \emph{average
end-to-end per-stream outage probability} defined as follows.
\begin{Def} [Average End-to-End Per-stream Outage Probability]
\label{Def:out_prob} Let $\mathcal H = \{\mathbf H_{SD}^{}, \mathbf
H_{SR}^{}, \mathbf H_{RD}^{}\}$ denote the \emph{aggregate channel
state} of the cooperative system. For source data stream
$\textbf{s}_{}^{(i)}$, let ${I\!\;}_{\textrm{PDF}}^{(i)} (\mathcal
H, \mathbf P_{\!S}^{}, \mathbf P_{\!R}^{})$ denote the end-to-end
mutual information given the channel state $\mathcal H$, the source
node precoder $\mathbf P_{\!S}^{}$, and the relay node precoder
$\mathbf P_{\!R}^{}$. Data stream $\textbf{s}_{}^{(i)}$ is in outage
if the end-to-end mutual information is less than the data rate of
$L/T$ bits per symbol, and the outage event can be modeled as
${I\!\;}_{\textrm{PDF}}^{(i)} (\mathcal H, \mathbf P_{\!S}^{},
\mathbf P_{\!R}^{}) < L/T$, $i = 1,2$. We define the average
end-to-end per-stream outage probability as
\begin{IEEEeqnarray}{l}
\textstyle P_{\textrm{out}}^{}( \mathbf P_{\!S}^{}, \mathbf
P_{\!R}^{} ) = \frac{1}{2} \big( \textrm{Pr} \, \big(
{I\!\;}_{\textrm{PDF}}^{(1)} (\mathcal H, \mathbf P_{\!S}^{},
\mathbf P_{\!R}^{}) \!<\! L/T \big) + \textrm{Pr} \, \big(
{I\!\;}_{\textrm{PDF}}^{(2)} (\mathcal H, \mathbf P_{\!S}^{},
\mathbf P_{\!R}^{}) \!<\! L/T \big)
\big).\;\;\;\;\label{eqn:def_out_prob}
\end{IEEEeqnarray}
~ \hfill\IEEEQEDclosed
\end{Def}

In consideration of efficient power utilization in the cooperative
system as well as limiting the total interference induced upon the
coverage area (which is usually restricted by government policy), we
seek to design the source node precoder $\mathbf P_{\!S}^{}$ and the
relay node precoder $\mathbf P_{\!R}^{}$ subject to a \emph{total
transmit power constraint} for the cooperative system\footnote{The
sum power constraint gives a first order constraint on the total
power resource of the network (accounting for the resources incurred
by adding more relays into the network). For instance, we can always
enhance performance by deploying more relays in the network and this
effect is not accounted for if we consider per node power
constraints only (cf. \cite{Jnl:SumPowerConstraint1,
Jnl:SumPowerConstraint2, Cnf:SumPowerConstraint}). The proposed
precoder design also readily accommodates per-node transmit power
constraints (cf. Remark~\ref{Rem:PerNodeTxPower}).}. Specifically,
the source node transmit power is given by $|| \mathbf P_{\!S}^{}
||_{}^{2}$, the relay node transmit power is given by $|| \mathbf
P_{\!R}^{} ||_{}^{2}$, and the total transmit power for the
cooperative system is given by $|| \mathbf P_{\!S}^{} ||_{}^{2} + ||
\mathbf P_{\!R}^{} ||_{}^{2}$.
\begin{problem}[Precoder Design for PDF with MMSE-SIC
Receiver]\label{Prob:MainProb} Let $P_{0}^{}$ denote the total
transmit power permitted for the cooperative system. The precoder
design problem to minimize the average end-to-end per-stream outage
probability is formulated as:
\begin{IEEEeqnarray*}{l}
\{ \mathbf P_{\!S}^{\star}, \mathbf P_{\!R}^{\star} \} =
\underset{\mathbf P_{\!S}^{}, \mathbf P_{\!R}^{}}{\arg \min}
\;P_{\textrm{out}}^{}( \mathbf P_{\!S}^{}, \mathbf P_{\!R}^{} ),
\;\;\textrm{s.t.}\;\; || \mathbf P_{\!S}^{} ||_{}^{2} + || \mathbf
P_{\!R}^{} ||_{}^{2} \leq
P_{0}^{}.\IEEEyesnumber\label{Eqn:OptProbMain}
\end{IEEEeqnarray*}
~ \hfill\IEEEQEDclosed
\end{problem}

Note that Problem~\ref{Prob:MainProb} is very difficult because it
is complicated to derive the closed-form expression for
$P_{\textrm{out}}^{}( \mathbf P_{\!S}^{}, \mathbf P_{\!R}^{} )$,
especially due to the non-linear SIC receiver structure. Since the
source and the relay nodes only have statistical CSI, we cannot
solve Problem~\ref{Prob:MainProb} by means of traditional channel
diagonalization schemes (e.g. \cite{Palomar03}). Moreover, it is
nontrivial to extend precoding schemes for point-to-point systems
with statistical CSIT and ML receiver \cite{Jnl:PrecoderStatCsit1,
Jnl:PrecoderStatCsit2} to the proposed problem, which requires
designing both source and relay node precoders matched to MMSE-SIC
receivers.

\section{MIMO Precoder Design for PDF Protocol and MMSE-SIC Receiver}
\label{sect:sol} In this section, we present the proposed precoder
design. First, we derive the closed-form expression for the average
end-to-end per-stream outage probability. Then, we employ a primal
decomposition approach \cite{Boyd03} to solve the precoder design
problem.

As shown in Section~\ref{Sec:PDF}, the PDF protocol results in three
cases (depending on which source data stream is correctly decoded
and forwarded by the relay node), and the average end-to-end outage
probability \eqref{eqn:def_out_prob} can be expressed as
\begin{IEEEeqnarray*}{ll}
P_{\textrm{out}}^{}( \mathbf P_{\!S}^{}, \mathbf P_{\!R}^{} ) \;&=\!
\textstyle\sum_{k=1}^{3} \frac{1}{2} \sum_{i=1}^{2} \textrm{Pr} \,
\big( \underbrace{ {I\!\;}_{\textrm{PDF}}^{(i)} (\mathcal H, \mathbf
P_{\!S}^{}, \mathbf P_{\!R}^{}) \!<\! L/T \,|\, \mathcal H \!\in\!
\mathcal H_{\!\mathcal{A}_{k}^{}}^{}}_{\textrm{Data stream }
\textbf{s}_{}^{(i)} \textrm{ is in outage under PDF~Case~}k.} \big),
\IEEEyesnumber\label{Eqn:AvgEnd2EndOutageProbI}
\end{IEEEeqnarray*}
where $\mathcal H_{\!\mathcal{A}_{k}^{}}^{}$ denotes all the
realizations of the aggregate channel state $\mathcal H$ that result
in PDF~Case~$k$. We denote the probability of PDF~Case~$k$ as
$\textrm{Pr} \, (\mathcal A_{k}^{})$ and the end-to-end outage
probability of data stream $\textbf{s}_{}^{(i)}$ under PDF~Case~$k$
as $P_{\textrm{out}}^{(i)} (\mathbf P_{\!S}^{}, \mathbf P_{\!R}^{}
\,|\, \mathcal A_{k}^{})$, so
\begin{IEEEeqnarray*}{l}
\textrm{Pr} \, \big( {I\!\;}_{\textrm{PDF}}^{(i)} (\mathcal H,
\mathbf P_{\!S}^{}, \mathbf P_{\!R}^{}) \!<\! L/T \,|\, \mathcal H
\!\in\! \mathcal H_{\!\mathcal{A}_{k}^{}}^{} \big) =
P_{\textrm{out}}^{(i)} (\mathbf P_{\!S}^{}, \mathbf P_{\!R}^{} \,|\,
\mathcal A_{k}^{}) \; \textrm{Pr} \, (\mathcal
A_{k}^{}).\IEEEyesnumber\label{Eqn:OutageProbPdfCase}
\end{IEEEeqnarray*}
It follows that \eqref{Eqn:AvgEnd2EndOutageProbI} can be
equivalently expressed as
\begin{IEEEeqnarray*}{l}
\textstyle P_{\textrm{out}}^{}( \mathbf P_{\!S}^{}, \mathbf
P_{\!R}^{} ) =\! \sum_{k=1}^{3} \frac{1}{2} \, \big(
P_{\textrm{out}}^{(1)} (\mathbf P_{\!S}^{}, \mathbf P_{\!R}^{} \,|\,
\mathcal A_{k}^{}) + P_{\textrm{out}}^{(2)} (\mathbf P_{\!S}^{},
\mathbf P_{\!R}^{} \,|\, \mathcal A_{k}^{}) \big) \; \textrm{Pr} \,
(\mathcal A_{k}^{}).\IEEEyesnumber\label{Eqn:AvgEnd2EndOutageProb}
\end{IEEEeqnarray*}

We summarize below in Lemma~\ref{lem:case_1} the probability of each
PDF case and the corresponding outage probability of each source
data stream. For analytical tractability, we assume that the source
and relay nodes employ \emph{orthogonal} STBC (OSTBC). In addition,
for notational convenience, we denote $D = 2_{}^{L/T} - 1$,
$\lambda_{SR}^{(1)} = || \mathbf H_{SR}^{} \, [\mathbf
P_{\!S}^{}]_{(:, 1:n_{S}^{(1)})}^{} ||_{}^{2}$, $\lambda_{SR}^{(2)}
= || \mathbf H_{SR}^{} \, [\mathbf P_{\!S}^{}]_{(:,
n_{S}^{(1)}+1:n_{\!S}^{})}^{} ||_{}^{2}$, $\lambda_{RD}^{} = ||
\mathbf H_{RD}^{} \, \mathbf P_{\!R}^{} ||_{}^{2}$,
$\lambda_{SD}^{(1)} = || \mathbf H_{SD}^{} \, [\mathbf
P_{\!S}^{}]_{(:, 1:n_{S}^{(1)})}^{} ||_{}^{2}$, and
$\lambda_{SD}^{(2)} = || \mathbf H_{SD}^{} \, [\mathbf
P_{\!S}^{}]_{(:, n_{S}^{(1)}+1:n_{\!S}^{})}^{} ||_{}^{2}$.
\begin{Lem}[Outage Probabilities under each PDF~Case]
\label{lem:case_1} The probability of PDF~Case~1 is
\begin{IEEEeqnarray*}{l}
\textstyle\textrm{Pr} \, (\mathcal A_{1}^{}) = \textrm{Pr} \bigg(\!
\frac{\lambda_{SR}^{(1)}}{\lambda_{SR}^{(2)} + N_{0}^{}} \ge D
\bigg),\IEEEyesnumber\label{eqn:ProbPdfCase1}
\end{IEEEeqnarray*}
and the end-to-end outage probabilities of data streams
$\textbf{s}_{}^{(1)}$ and $\textbf{s}_{}^{(2)}$ are given by
\begin{IEEEeqnarray*}{l}
\textstyle P_{\textrm{out}}^{(1)} (\mathbf P_{\!S}^{}, \mathbf
P_{\!R}^{} |\, \mathcal A_{1}^{}) \!=\! \textrm{Pr} \Bigg(\!
\frac{\big( (\lambda_{SD}^{(1)})_{}^{1\!/2} +
(\lambda_{RD}^{})_{}^{1\!/2} \big)_{}^{2}}{\lambda_{SD}^{(2)} +
2N_{0}^{}} < D \Bigg)\IEEEyesnumber\label{eqn:OutPdfCase1}\\
\textstyle P_{\textrm{out}}^{(2)} (\mathbf P_{\!S}^{}, \mathbf
P_{\!R}^{} |\, \mathcal A_{1}^{}) \!=\! ( 1 \!-\!
P_{\textrm{out}}^{(1)} (\mathbf P_{\!S}^{}, \mathbf P_{\!R}^{} |\,
\mathcal A_{1}^{}) ) \, \textrm{Pr} \, ( \lambda_{SD}^{(2)} \!<\! D
) \!+\! P_{\textrm{out}}^{(1)} (\mathbf P_{\!S}^{}, \mathbf
P_{\!R}^{} |\, \mathcal A_{1}^{}) \, \textrm{Pr} \bigg(\!
\frac{\lambda_{SD}^{(2)}}{4 \, \epsilon_{\!\mathcal{A}_{1}^{}}^{} \!
\lambda_{SD}^{(1)} + N_{0}^{}} \!<\! D \bigg)
\end{IEEEeqnarray*}
where $\epsilon_{\mathcal{A}_{1}^{}}^{}\!$ denotes the symbol error
rate (SER) in decoding the cooperative stream.
The probability of PDF~Case~2 is
\begin{IEEEeqnarray*}{l}
\textstyle\textrm{Pr} \, (\mathcal A_{2}^{}) = \textrm{Pr} \bigg(\!
\frac{\lambda_{SR}^{(1)}}{\lambda_{SR}^{(2)} + N_{0}^{}} < D \bigg)
\, \textrm{Pr} \bigg(\! \frac{\lambda_{SR}^{(2)}}{4
\epsilon_{\!R}^{} \lambda_{SR}^{(1)} + N_{0}^{}} \ge D
\bigg)\IEEEyesnumber\label{eqn:ProbPdfCase2}
\end{IEEEeqnarray*}
where $\epsilon_{R}^{}$ denotes the SER in decoding data stream
$\textbf{s}_{}^{(1)}$ by the relay node, and the end-to-end outage
probabilities of data streams $\textbf{s}_{}^{(1)}$ and
$\textbf{s}_{}^{(2)}$ are given by
\begin{IEEEeqnarray*}{l}
\textstyle P_{\textrm{out}}^{(1)} (\mathbf P_{\!S}^{}, \mathbf
P_{\!R}^{} |\, \mathcal A_{2}^{}) \!=\! ( 1 \!-\!
P_{\textrm{out}}^{(2)} (\mathbf P_{\!S}^{}, \mathbf P_{\!R}^{} |\,
\mathcal A_{2}^{}) ) \, \textrm{Pr} \, ( \lambda_{SD}^{(1)} \!<\! D
) \!+\! P_{\textrm{out}}^{(2)} (\mathbf P_{\!S}^{}, \mathbf
P_{\!R}^{} |\, \mathcal A_{2}^{}) \, \textrm{Pr} \bigg(\!
\frac{\lambda_{SD}^{(1)}}{4 \, \epsilon_{\!\mathcal{A}_{2}^{}}^{} \!
\lambda_{SD}^{(2)} + N_{0}^{}} \!<\! D
\bigg)\\
\textstyle P_{\textrm{out}}^{(2)} (\mathbf P_{\!S}^{}, \mathbf
P_{\!R}^{} |\, \mathcal A_{2}^{}) \!=\! \textrm{Pr} \Bigg(\!
\frac{\big( (\lambda_{SD}^{(2)})_{}^{1\!/2} +
(\lambda_{RD}^{})_{}^{1\!/2} \big)_{}^{2}}{\lambda_{SD}^{(1)} +
2N_{0}^{}} < D \Bigg)\IEEEyesnumber\label{eqn:OutPdfCase2}
\end{IEEEeqnarray*}
where $\epsilon_{\mathcal{A}_{2}^{}}^{}$ denotes the SER in decoding
the cooperative stream.
\hbox{The probability of PDF~Case~3 is}
\begin{IEEEeqnarray*}{l}
\textstyle\textrm{Pr} \, (\mathcal A_{3}^{}) = \textrm{Pr} \bigg(\!
\frac{\lambda_{SR}^{(1)}}{\lambda_{SR}^{(2)} + N_{0}^{}} < D \bigg)
\, \textrm{Pr} \bigg(\! \frac{\lambda_{SR}^{(2)}}{4
\epsilon_{\!R}^{} \lambda_{SR}^{(1)} + N_{0}^{}} < D
\bigg)\IEEEyesnumber\label{eqn:ProbPdfCase3}
\end{IEEEeqnarray*}
where $\epsilon_{R}^{}$ denotes the SER in decoding data stream
$\textbf{s}_{}^{(1)}$ by the relay node, and the end-to-end outage
probabilities of data streams $\textbf{s}_{}^{(1)}$ and
$\textbf{s}_{}^{(2)}$ are given by
\begin{IEEEeqnarray*}{ll}
\textstyle P_{\textrm{out}}^{(1)} (\mathbf P_{\!S}^{}, \mathbf
P_{\!R}^{} |\, \mathcal A_{3}^{}) \!=\! \textrm{Pr} \bigg(\!
\frac{\lambda_{SD}^{(1)}}{\lambda_{SD}^{(2)} + N_{0}^{}} <
D \bigg)\IEEEyesnumber\label{eqn:OutPdfCase3}\\
\textstyle P_{\textrm{out}}^{(2)} (\mathbf P_{\!S}^{}, \mathbf
P_{\!R}^{} |\, \mathcal A_{3}^{}) \!=\! ( 1 \!-\!
P_{\textrm{out}}^{(1)} (\mathbf P_{\!S}^{}, \mathbf P_{\!R}^{} |\,
\mathcal A_{3}^{}) ) \, \textrm{Pr} \, ( \lambda_{SD}^{(2)} \!<\! D
) \!+\! P_{\textrm{out}}^{(1)} (\mathbf P_{\!S}^{}, \mathbf
P_{\!R}^{} |\, \mathcal A_{3}^{}) \, \textrm{Pr} \bigg(\!
\frac{\lambda_{SD}^{(2)}}{4 \epsilon_{\!D}^{} \lambda_{SD}^{(1)} +
N_{0}^{}} \!<\! D \bigg)
\end{IEEEeqnarray*}
where $\epsilon_{D}^{}$ denotes the SER in decoding
$\textbf{s}_{}^{(1)}$ by the destination node.
\end{Lem}
\proof Please refer to Appendix~A for the proof.
\endproof
The average end-to-end per-stream outage probability
$P_{\textrm{out}}^{}( \mathbf P_{\!S}^{}, \mathbf P_{\!R}^{} )$ can
be deduced from
\eqref{Eqn:AvgEnd2EndOutageProb}--\eqref{eqn:OutPdfCase3}, but it is
non-trivial to design the precoders $\mathbf P_{\!S}^{}$ and
$\mathbf P_{\!R}^{}$ to minimize $P_{\textrm{out}}^{}( \mathbf
P_{\!S}^{}, \mathbf P_{\!R}^{} )$. In the following, we solve the
precoder design problem using a primal decomposition approach
\cite{Boyd03}, where we tailor to the characteristics of the PDF
protocol to derive efficient precoder solutions.
\begin{problem}[Precoder Design Based on Primal
Decomposition]\label{Prob:PD} We introduce the auxiliary variables
$\alpha_{S}^{}$ and $\alpha_{R}^{}$. The precoder design problem can
be decomposed into the subproblems and master problem below.
\begin{IEEEeqnarray*}{l}
\textrm{\emph{Subproblem 1 (Optimization w.r.t. }$\mathbf
P_{\!R}^{}$\emph{):}}\;\, P_{\textrm{out}}^{\star} (\mathbf
P_{\!S}^{}, \alpha_{R}^{}) \!=\! \min_{\mathbf P_{\!R}^{}}
P_{\textrm{out}}^{}( \mathbf P_{\!S}^{}, \mathbf P_{\!R}^{} ),
\;\,\textrm{s.t.}\;\, || \mathbf P_{\!R}^{} ||_{}^{2} \!\leq\!
\alpha_{R}^{}.\;\;\;\;\;\;\IEEEyesnumber\label{Eqn:OptProbSubP_R}
\end{IEEEeqnarray*}
\begin{IEEEeqnarray*}{l}
\textrm{\emph{Subproblem 2 (Optimization w.r.t. }$\mathbf
P_{\!S}^{}$\emph{):}}\;\, P_{\textrm{out}}^{\star} (\alpha_{S}^{},
\alpha_{R}^{}) \!=\! \min_{\mathbf P_{\!S}^{}}
P_{\textrm{out}}^{\star} (\mathbf P_{\!S}^{}, \alpha_{R}^{}),
\;\,\textrm{s.t.}\;\, || \mathbf P_{\!S}^{} ||_{}^{2} \!\leq\!
\alpha_{S}^{}.\;\;\;\;\;\;\IEEEyesnumber\label{Eqn:OptProbSubP_S}
\end{IEEEeqnarray*}
\begin{IEEEeqnarray*}{l}
\textrm{\emph{Master Problem (Optimization w.r.t. }$\alpha_{R}^{},
\alpha_{S}^{}$\emph{):}}\;\, P_{\textrm{out}}^{\star} \!=\!\!
\min_{\alpha_{S}^{}, \alpha_{R}^{} \;\ge 0}\!
P_{\textrm{out}}^{\star} (\alpha_{S}^{}, \alpha_{R}^{}),
\;\,\textrm{s.t.}\;\, \alpha_{S}^{} \!+\! \alpha_{R}^{} \!\leq\!
P_{0}^{}.\;\;\;\;\;\;\IEEEyesnumber\label{Eqn:OptProbMaster}
\end{IEEEeqnarray*}
~ \hfill\IEEEQEDclosed
\end{problem}

The master problem \eqref{Eqn:OptProbMaster} determines the
\emph{power budget allocation} with respect to (w.r.t.) the total
transmit power constraint, where the relative values of
$\alpha_{S}^{}$ and $\alpha_{R}^{}$ affect the outage probabilities
of transmission from the source node and from the relay node. For a
given tuple of $\alpha_{R}^{}$ and $\alpha_{S}^{}$, we solve
subproblems \eqref{Eqn:OptProbSubP_R} and \eqref{Eqn:OptProbSubP_S}
to derive the particular precoder solutions\footnote{It is not
analytically tractable to find global optimal solutions so we apply
some practically motivated approximations to obtain efficient
suboptimal solutions.}.

\begin{Rem}[Accommodating Per-node Transmit Power Constraints]
\label{Rem:PerNodeTxPower} The proposed precoder design can readily
accommodate per-node transmit power constraints. For such settings,
we directly set $\alpha_{S}^{}$ and $\alpha_{R}^{}$ equal to the
source node and relay node transmit power constraints, respectively,
and solve subproblems \eqref{Eqn:OptProbSubP_R} and
\eqref{Eqn:OptProbSubP_S} to derive the precoder solutions.~
\hfill\IEEEQEDclosed
\end{Rem}

First, the relay node precoder design is given by the following
theorem.
\begin{Thm}[Relay Precoder Design]
\label{Thm:relay}The relay node precoder solution to subproblem
\eqref{Eqn:OptProbSubP_R} is given by
\begin{IEEEeqnarray}{l}
\textstyle\mathbf P_{\!R}^{\star} =
\sqrt{\frac{\alpha_{R}^{}}{\textrm{Tr} \,(( \mathbf
\Sigma_{\!RD,t}^{} )_{}^{-1})}} \, ( \mathbf U_{\!RD,t}^{}
)_{}^{\ast} \, ( \mathbf \Sigma_{\!RD,t}^{} )_{}^{-1/2}.
\end{IEEEeqnarray}
$\mathbf U_{\!RD,t}^{}$ and $\mathbf \Sigma_{\!RD,t}^{}$ are given
by the eigendecomposition $( \mathbf \Lambda_{RD,t}^{}
)_{}^{\!1\!\!\:/\!\:2} ( ( \mathbf \Lambda_{RD,t}^{}
)_{}^{\!1\!\!\:/\!\:2} )_{}^{\dag} = \mathbf U_{\!RD,t}^{} \mathbf
\Sigma_{\!RD,t}^{} ( \mathbf U_{\!RD,t}^{} )_{}^{\dag}$, where
$\mathbf \Lambda_{RD,t}^{}$ is the RD link transmit-side correlation
matrix.
\end{Thm}
\proof Please refer to Appendix~B for the proof.
\endproof

The average end-to-end per-stream outage probability given the relay
node precoder $\mathbf P_{\!R}^{\star}$ is
\begin{IEEEeqnarray*}{l}
\textstyle P_{\textrm{out}}^{\star} (\mathbf P_{\!S}^{},
\alpha_{R}^{}) =\! \sum_{k=1}^{3} \frac{1}{2} \, \big(
P_{\textrm{out}}^{(1)} (\mathbf P_{\!S}^{}, \mathbf P_{\!R}^{\star}
\,|\, \mathcal A_{k}^{}) + P_{\textrm{out}}^{(2)} (\mathbf
P_{\!S}^{}, \mathbf P_{\!R}^{\star} \,|\, \mathcal A_{k}^{}) \big)
\; \textrm{Pr} \, (\mathcal
A_{k}^{}).\IEEEyesnumber\label{Eqn:OutageGivenP_R_Star}
\end{IEEEeqnarray*}
As discussed in Remark~\ref{Rem:RemarkLinkConditions}, in general,
the relay link (i.e. transmission through the SR and RD links) is
much stronger than the direct link (i.e. transmission through the SD
link). Comparing \eqref{eqn:OutPdfCase1}, \eqref{eqn:OutPdfCase2},
\eqref{eqn:OutPdfCase3}, it is straightforward to show that
\begin{IEEEeqnarray*}{l}
P_{\textrm{out}}^{(1)} (\mathbf P_{\!S}^{}, \mathbf P_{\!R}^{\star}
\,|\, \mathcal A_{3}^{}) \gg P_{\textrm{out}}^{(1)} (\mathbf
P_{\!S}^{}, \mathbf P_{\!R}^{\star} \,|\, \mathcal A_{1}^{}),
P_{\textrm{out}}^{(2)} (\mathbf P_{\!S}^{}, \mathbf P_{\!R}^{\star}
\,|\, \mathcal A_{2}^{})\\
P_{\textrm{out}}^{(2)} (\mathbf P_{\!S}^{}, \mathbf P_{\!R}^{\star}
\,|\, \mathcal A_{3}^{}) \gg P_{\textrm{out}}^{(2)} (\mathbf
P_{\!S}^{}, \mathbf P_{\!R}^{\star} \,|\, \mathcal A_{1}^{}),
P_{\textrm{out}}^{(1)} (\mathbf P_{\!S}^{}, \mathbf P_{\!R}^{\star}
\,|\, \mathcal A_{2}^{})\IEEEyesnumber\label{Eqn:xxx}
\end{IEEEeqnarray*}
so $P_{\textrm{out}}^{\star} (\mathbf P_{\!S}^{}, \alpha_{R}^{})$ is
in fact \emph{dominated} by the outage probabilities under
PDF~Case~3. Therefore, we should seek to design the source node
precoder $\mathbf P_{\!S}^{}$ to minimize the probability of
PDF~Case~3, $\textrm{Pr} \, (\mathcal A_{3}^{})$. In effect, we are
interested to design the source node precoder $\mathbf P_{\!S}^{}$
to best improve the reliability of the \emph{already-strong} relay
link, and thereby provide a high-quality cooperative stream to the
destination node for MMSE-SIC receiver to avoid error propagation.
The source node precoder design is given by the following theorem.
\begin{Thm}[Source Precoder Design]
\label{Thm:source} The source node precoder solution to subproblem
\eqref{Eqn:OptProbSubP_S} is given by
\begin{IEEEeqnarray*}{l}
\mathbf P_{\!S}^{\star} = \sqrt{\alpha_{S}^{}} \, ( \mathbf
U_{\!SR,t}^{} )_{}^{\ast} \! \left[\!\!
\begin{array}{cc}\sqrt{\frac{\rho_{S}^{(1)}}{\textrm{Tr} \,((
\mathbf \Sigma_{\!SR,t}^{(1)} )_{}^{-1})}} ( \mathbf
\Sigma_{\!SR,t}^{(1)})_{}^{-1/2}\!\!\!\!\!\!& \textbf{0}\\
\textbf{0}\!\!\!\!\!\!& \sqrt{\frac{\rho_{S}^{(2)}}{\textrm{Tr} \,((
\mathbf \Sigma_{\!SR,t}^{(2)} )_{}^{-1})}} ( \mathbf
\Sigma_{\!SR,t}^{(2)})_{}^{-1/2}\end{array}\!\!\right].\IEEEyesnumber\label{Eqn:P_SStructure}
\end{IEEEeqnarray*}
$\mathbf U_{\!SR,t}^{}$, $\mathbf \Sigma_{\!SR,t}^{(1)}$ = $[\mathbf
\Sigma_{\!SR,t}^{}]_{(1:n_{S}^{(1)},1:n_{S}^{(1)})}^{}$, and
$\mathbf \Sigma_{\!SR,t}^{(2)}$ = $[\mathbf
\Sigma_{\!SR,t}^{}]_{(n_{S}^{(1)}+1:n_{\!S}^{},
n_{S}^{(1)}+1:n_{\!S}^{})}^{}$ are given by the eigendecomposition
$( \mathbf \Lambda_{SR,t}^{} )_{}^{\!1\!\!\:/\!\:2} ( ( \mathbf
\Lambda_{SR,t}^{} )_{}^{\!1\!\!\:/\!\:2} )_{}^{\dag} = \mathbf
U_{\!SR,t}^{} \mathbf \Sigma_{\!SR,t}^{} ( \mathbf U_{\!SR,t}^{}
)_{}^{\dag}$, where $\mathbf \Lambda_{SR,t}^{}$ is the SR link
transmit-side correlation matrix. $\rho_{S}^{(1)}$ and
$\rho_{S}^{(2)}$ are per-stream power allocation variables with
$\rho_{S}^{(1)} + \rho_{S}^{(2)} = 1$.
\end{Thm}
\proof Please refer to Appendix~C for the proof.
\endproof

The master problem \eqref{Eqn:OptProbMaster} belongs to the class of
quasi-convex optimization problem \cite{Boyd03} and can be
efficiently solved using, for example, bisection search algorithms.
In essence, the master problem determines $\alpha_{R}^{}$ and
$\alpha_{S}^{}$ that control how much we rely on the relay link for
partial forwarding. If the SR and RD links are in good condition, we
allocate more power to the relay link and increase $\alpha_{R}^{}$;
otherwise, we allocate more power to the direct link and increase
$\alpha_{S}^{}$.

For the system under study, Problem~\ref{Prob:PD} can be solved in a
distributed fashion. As per Assumption~\ref{Assumption:Csi}, since
the source node has statistical CSI of all links, it can solve
Problem~\ref{Prob:PD} and feed back the scalar $\alpha_{R}^{}$ to
the relay node. In turn, the relay node, which only has statistical
CSI of the RD link, can \emph{locally} solve subproblem
\eqref{Eqn:OptProbSubP_R} to design the relay node precoder.

\section{Simulation Results}
\label{sect:experiments} 
\begin{figure}[!ht]
\centering \subfloat[][]{\includegraphics[width = 5in]{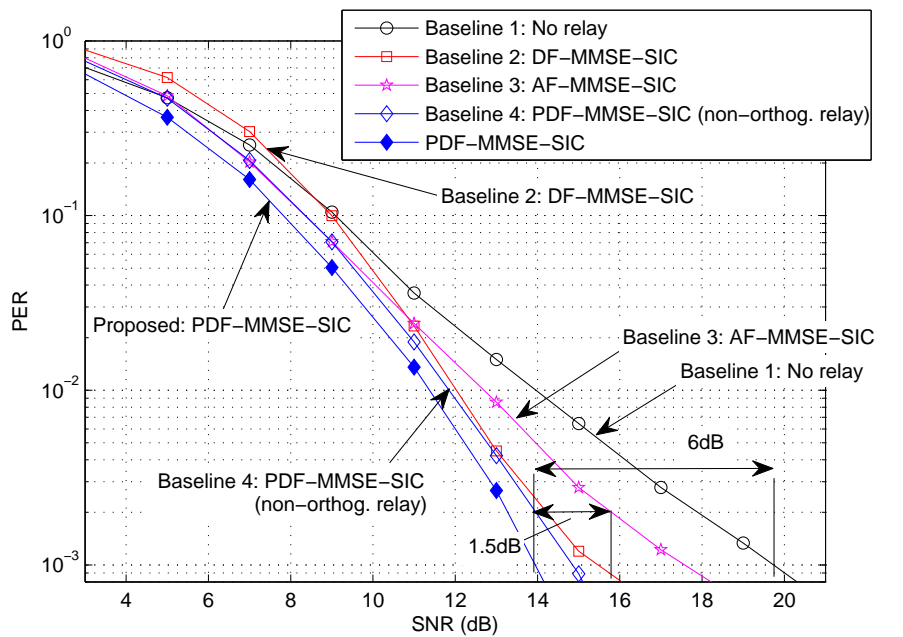}
\label{fig:PerWithPrecodingQpsk}}\\
\subfloat[][]{\includegraphics[width =
5in]{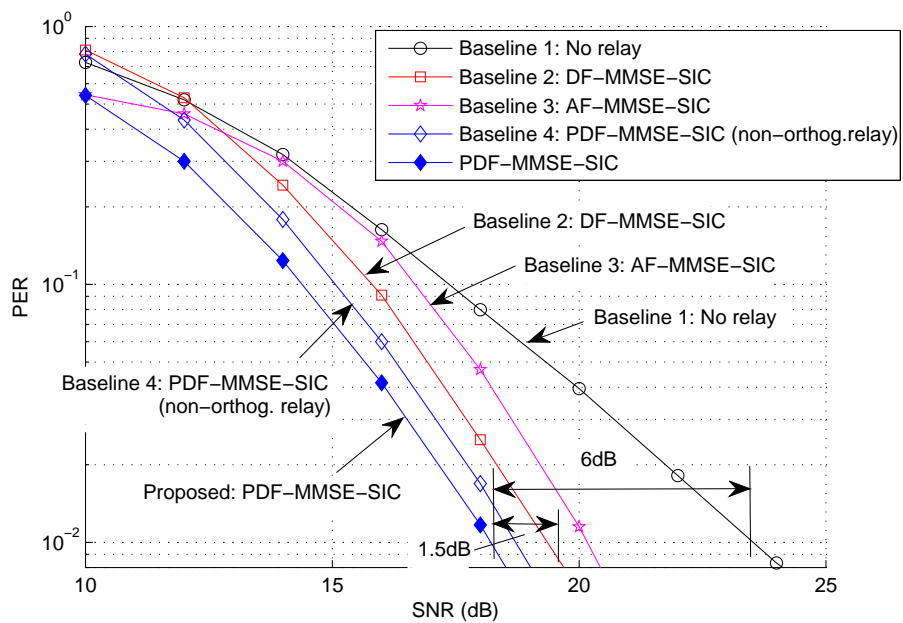}\label{fig:PerWithPrecoding16Qam}}
\caption{PER versus receive SNR comparison between the proposed
PDF-MMSE-SIC scheme and baseline schemes. The source node is
equipped with 4 antennas, whereas the relay and destination nodes
are equipped with 2 antennas. We adopt the path loss models
specified in Fig.~\ref{fig:link_budget}: the source, relay and
destination nodes are located according to the topology in
Fig.~\ref{fig:link_budget} with $d_{SR}^{} = 400~\textrm{m}$,
$d_{RD}^{} = 300~\textrm{m}$, and $d_{SD}^{} = 500~\textrm{m}$.
(a)~Performance with QPSK modulation. (b)~Performance with 16-QAM
modulation.} \label{fig:PerWithPrecoding}
\end{figure}

\begin{figure}
\centering
\includegraphics[width = 5in]{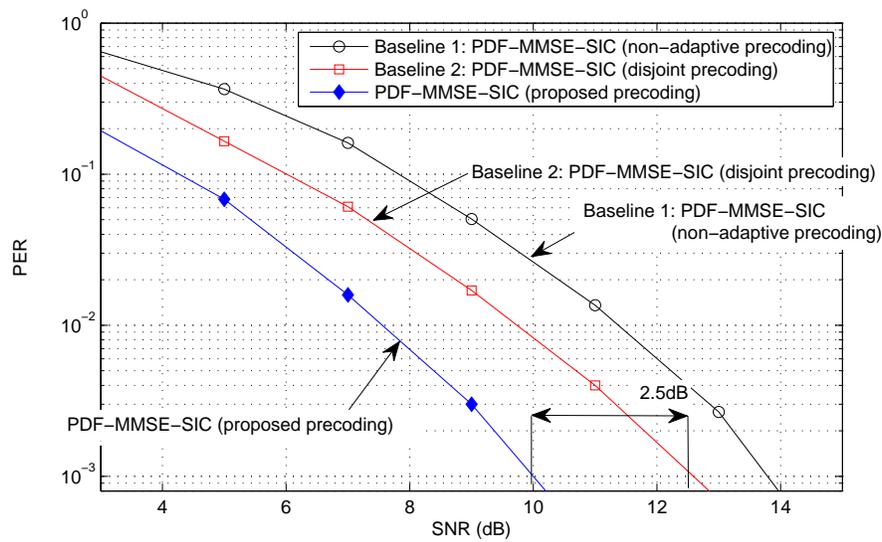}
\caption{PER versus receive SNR comparison between the proposed
precoding structure and baseline schemes. The source node is
equipped with 4 antennas, whereas the relay and destination nodes
are equipped with 2 antennas. The data streams are modulated using
QPSK. We adopt the path loss models specified in
Fig.~\ref{fig:link_budget}: the source, relay and destination nodes
are located according to the topology in Fig.~\ref{fig:link_budget}
with $d_{SR}^{} = 400~\textrm{m}$, $d_{RD}^{} = 300~\textrm{m}$, and
$d_{SD}^{} = 500~\textrm{m}$.} \label{fig:PerPdfCompare}
\end{figure}

In this section, we provide numerical
simulation results to assess the performance of the proposed PDF
protocol (cf. Fig.~\ref{fig:PerWithPrecoding}) and MIMO precoder
design (cf. Fig.~\ref{fig:PerPdfCompare}).

For the purpose of illustration, we consider the following practical
multi-antenna cooperative system with settings similar to those
defined in the IEEE~802.16m standard \cite{Std:16m}. We assume that
uniform linear antenna arrays are used \cite{Std:4G-EMD:16m}, where
the source node has $n_{S}^{}= 4$ antennas, the relay node has
$n_{R}^{} = 2$ antennas, and the destination node has $n_{D}^{} = 2$
antennas. At the source node, the data streams $\mathbf s_{}^{(1)}$
and $\mathbf s_{}^{(2)}$ are transmitted through
$n_{S}^{(1)}=n_{S}^{(2)} = 2$ diversity streams. We assume the
source, the relay, and the destination nodes are located according
to the topology in Fig.~\ref{fig:link_budget} with distances
$d_{SR}^{} = 400~\textrm{m}$, $d_{RD}^{} = 300~\textrm{m}$, and
$d_{SD}^{} = 500~\textrm{m}$. We evaluate performance using the
end-to-end packet error rate (PER) versus SNR as metric.
Specifically, we encode the data streams using the convolutional
turbo code defined in the IEEE~802.16m standard
\cite[Section~16.3.11.1.5]{Std:16m}: we assume each transmission
phase lasts for $T=96$ symbol time slots, where each data stream
contains $L=12$ information bytes coded at rate $1/2$ and modulated
using QPSK or 16-QAM.

We show in Fig.~\ref{fig:PerWithPrecoding} the performance of the
proposed PDF protocol with non-adaptive precoding\footnote{The
source and relay node precoders are given by random unitary matrices
and the total transmit power is evenly allocated between the source
and relay nodes.}. We compare the proposed PDF-MMSE-SIC relay
protocol against the following baselines.
\begin{list}{\labelitemi}{\leftmargin=0.5em}
\item Baseline 1 (No relay): A relay node is not deployed and the
destination node can only receive from the direct SD link, where the
destination node uses MMSE-SIC receiver.
\item Baseline 2 (DF MMSE-SIC): The relay node adopts the DF
protocol (cf. \eqref{Eqn:DfRelaySymbol}), where the relay and
destination nodes use MMSE-SIC receivers.
\item Baseline 3 (AF MMSE-SIC): The relay node adopts the AF
protocol (cf. \eqref{Eqn:AfRelaySymbol}), where the relay and
destination nodes use MMSE-SIC receivers.
\item Baseline 4 (PDF MMSE-SIC with non-orthogonal relaying): The source
node repeats transmission in both listening and cooperative phases
\cite{Non-OrthogonalRelay}, where the relay and destination nodes
use MMSE-SIC receivers.
\end{list}
Let us focus on the performance with QPSK modulation (cf.
\figurename~\ref{fig:PerWithPrecodingQpsk}). It can be seen that at
PER of $10_{}^{-3}$ the PDF-MMSE-SIC protocol has SNR gain in excess
of 6~dB compared to when a relay node is not deployed, and has SNR
gains of over 1.5~dB compared to conventional DF-MMSE-SIC and
AF-MMSE-SIC schemes. The superior error performance by applying the
PDF protocol is manifested from more effective mitigation of
inter-stream interference at the destination node (compared to
DF-MMSE-SIC and AF-MMSE-SIC) as well as enhanced probability that
the relay node can assist with data transmission (compared to
DF-MMSE-SIC). Note that it is inefficient to perform non-orthogonal
relaying since transmission by the source node in the cooperative
phase increases inter-stream interference in the decoding process at
the destination node.

We demonstrate in Fig.~\ref{fig:PerPdfCompare} the effectiveness of
the proposed precoding structure by comparing it with the following
baselines:
\begin{list}{\labelitemi}{\leftmargin=0.5em}
\item Baseline 1 (PDF-MMSE-SIC with non-adaptive precoding): The
basic PDF protocol.
\item Baseline 2 (PDF-MMSE-SIC with disjoint precoding): The source
relay node precoders are determined in similar fashion as
Theorem~\ref{Thm:relay} and \ref{Thm:source} but the total transmit
power is evenly allocated between the source and relay nodes.
\end{list}
It can be seen that the proposed precoding structure yields better
error performance than baselines 1 and 2 for all SNR regime. For
instance, at PER of $10_{}^{-3}$ the proposed precoder design has
over 4 dB SNR gain over non-adaptive precoding (Baseline 1).
Compared to disjoint precoding (Baseline 2), the proposed design
achieves substantial advantage by adapting the power constraints of
the source and relay nodes.

\section{Conclusions}
\label{sect:conclusion} In this paper, we consider precoder design
at the source and relay nodes for correlated multi-antenna
cooperative systems that are matched to the PDF relay protocol and
MMSE-SIC receivers. We derived the closed-form solution of the
precoders at the source and relay nodes based on a primal
decomposition approach. The performance of the proposed precoder
designs is compared with several baselines and is shown to achieve
significant performance gain compared to the baseline systems with
MMSE-SIC receiver.

\section*{Appendix A: Proof of Lemma~\ref{lem:case_1}}
\label{pf:case_1} We first show how to derive the mutual information
for the SR link; in a similar fashion we can derive the mutual
information for each PDF case. As per
\eqref{Eqn:ListeningPhaseReceivedSignalsR}, the source data streams
$\textbf{s}_{}^{(1)}$ and $\textbf{s}_{}^{(2)}$ are respectively
encoded into $\textbf{X}_{}^{(1)}$ and $\textbf{X}_{}^{(2)}$, and
the received signals of the relay node are given by
\begin{IEEEeqnarray*}{l}
\mathbf Y_{\!R,L}^{} = \mathbf H_{SR}^{} \, [\mathbf
P_{\!S}^{}]_{(:, 1:n_{S}^{(1)})}^{} \mathbf X_{}^{(1)} + \mathbf
H_{SR}^{} \, [\mathbf P_{\!S}^{}]_{(:, n_{S}^{(1)}+1:n_{\!S}^{})}^{}
\mathbf X_{}^{(2)} \!+\; \mathbf
Z_{\!R,L}^{}.\IEEEyesnumber\label{eqn:PerStreamMIModel1}
\end{IEEEeqnarray*}
Suppose we decode $\textbf{s}_{}^{(1)}$ first while treating
$\textbf{s}_{}^{(2)}$ as interference. Let $\lambda_{SR}^{(1)} = ||
\mathbf H_{SR}^{} \, [\mathbf P_{\!S}^{}]_{(:, 1:n_{S}^{(1)})}^{}
||_{}^{2}$ and $\lambda_{SR}^{(2)} = || \mathbf H_{SR}^{} \,
[\mathbf P_{\!S}^{}]_{(:, n_{S}^{(1)}+1:n_{\!S}^{})}^{} ||_{}^{2}$.
As shown in \cite{Ganesan01}, after space-time processing the
\emph{effective} signal model for $\textbf{s}_{}^{(1)}$ is
$\textbf{y}_{\!R,L}^{(1)} = \lambda_{SR}^{(1)}\, \textbf{s}_{}^{(1)}
+ \textbf{z}_{\!R,L}^{(1)}$, where $\textbf{z}_{\!R,L}^{(1)}$
denotes white Gaussian aggregate interference and noise terms with
zero mean
and variance
$\lambda_{SR}^{(1)} \lambda_{SR}^{(2)} + \lambda_{SR}^{(1)}
N_{0}^{}$. Specifically, the noise variance is given by $|| (
\mathbf H_{SR}^{} \, [\mathbf P_{\!S}^{}]_{(:, 1:n_{S}^{(1)})}^{}
)_{}^{\dag} \; \mathbf H_{SR}^{} \, [\mathbf P_{\!S}^{}]_{(:,
n_{S}^{(1)}+1:n_{\!S}^{})}^{} ||_{}^{2} + \lambda_{SR}^{(1)}
N_{0}^{} \leq \lambda_{SR}^{(1)} \lambda_{SR}^{(2)} +
\lambda_{SR}^{(1)} N_{0}^{}$. The SINR of $\textbf{s}_{}^{(1)}$ is
given by $\gamma_{\!R,L}^{(1)} =
\frac{\lambda_{SR}^{(1)}}{\lambda_{SR}^{(2)} + N_{0}^{}}$. Suppose
$\textbf{s}_{}^{(1)}$ is decoded as
$\widehat{\textbf{s}}{\,}_{R}^{(1)}$ at SER $\epsilon_{R}^{}$. We
re-encode $\widehat{\textbf{s}}{\,}_{R}^{(1)}$ into
$\widehat{\textbf{X}}{}_{}^{(1)}$ and cancel it from the received
signals $\mathbf Y_{\!R,L}^{}$; the resultant signals are
\begin{IEEEeqnarray*}{l}
\widetilde{\textbf{Y}}{}_{\!R,L}^{} = \mathbf H_{SR}^{} \, [\mathbf
P_{\!S}^{}]_{(:, n_{S}^{(1)}+1:n_{\!S}^{})}^{} \mathbf X_{}^{(2)} +
\mathbf H_{SR}^{} \, [\mathbf P_{\!S}^{}]_{(:, 1:n_{S}^{(1)})}^{} (
\mathbf X_{}^{(1)} - \widehat{\mathbf X}{}_{}^{(1)} ) +\; \mathbf
Z_{\!R,L}^{}.
\IEEEyesnumber\IEEEyesnumber\label{eqn:PerStreamMIModel2}
\end{IEEEeqnarray*}
After space-time processing, we can express the \emph{effective}
signal model for data stream $\textbf{s}_{}^{(2)}$ as
$\textbf{y}_{\!R,L}^{(2)} = \lambda_{SR}^{(2)} \,
\textbf{s}_{}^{(2)} + \textbf{z}_{}^{(2)}$, where
$\textbf{z}_{}^{(2)}$ denotes zero-mean white Gaussian 
aggregate residual interference and noise terms. The variance of
$\textbf{z}_{}^{(2)}$ depends on whether data stream
$\textbf{s}_{}^{(1)}$ is correctly decoded: if
$\textbf{s}_{}^{(1)} = \widehat{\textbf{s}}{\,}_{R}^{(1)}$, 
the variance of $\textbf{z}_{}^{(2)}$ is given by
$\lambda_{SR}^{(2)}N_{0}^{}$; otherwise, the variance of
$\textbf{z}_{}^{(2)}$ is $4 \epsilon_{R}^{} \lambda_{SR}^{(2)}
\lambda_{SR}^{(1)} + \lambda_{SR}^{(2)}N_{0}^{}$. Specifically, the
noise variance is given by\\\centerline{$|| ( \mathbf H_{SR}^{} \,
[\mathbf P_{\!S}^{}]_{(:, n_{S}^{(1)}+1:n_{\!S}^{})}^{} )_{}^{\dag}
\; \mathbf H_{SR}^{} \, [\mathbf P_{\!S}^{}]_{(:, 1:n_{S}^{(1)})}^{}
||_{}^{2} \mathbb{E} ( || \textbf{X}_{}^{(1)} -
\widehat{\textbf{X}}{}_{}^{(1)} ||_{}^{2} ) +
\lambda_{SR}^{(2)}N_{0}^{}$.} If $\textbf{X}_{}^{(1)} \!\neq\!
\widehat{\textbf{X}}{}_{}^{(1)}$, $\mathbb{E} ( ||
\textbf{X}_{}^{(1)} \!-\! \widehat{\textbf{X}}{}_{}^{(1)} ||_{}^{2}
) \!\leq\! \mathbb{E} ( || \textbf{X}_{}^{(1)} ||_{}^{2} ) \!+\!
\mathbb{E} ( 2 \textrm{Re} ( \textbf{X}_{}^{(1)} (
\widehat{\textbf{X}}{}_{}^{(1)} )_{}^{\dag} ) ) \!+\! \mathbb{E} (
|| \widehat{\textbf{X}}{}_{}^{(1)} ||_{}^{2} ) \!\leq\! 4 \mathbb{E}
( || \textbf{X}_{}^{(1)} ||_{}^{2} )$. Hence, $\mathbb{E} ( ||
\textbf{X}_{}^{(1)} - \widehat{\textbf{X}}{}_{}^{(1)} ||_{}^{2} )
\leq 4 \epsilon_{R}^{}$, and the noise variance can be expressed as
$4 \epsilon_{R}^{} \lambda_{SR}^{(2)} \lambda_{SR}^{(1)} +
\lambda_{SR}^{(2)}N_{0}^{}$.
Correspondingly, if $\textbf{s}_{}^{(1)} =
\widehat{\textbf{s}}{\,}_{R}^{(1)}$, the SINR of
$\textbf{s}_{}^{(2)}$ is given by $\gamma_{\!R,L}^{(2)} =
\lambda_{SR}^{(2)}\,/\!N_{0}^{}$; otherwise, the SINR of
$\textbf{s}_{}^{(2)}$ is given by $\gamma_{\!R,L}^{(2)} =
\frac{\lambda_{SR}^{(2)}}{4 \epsilon_{\!R}^{} \lambda_{SR}^{(1)} +
N_{0}^{}}$. 
Therefore, the mutual information for $\textbf{s}_{}^{(1)}$ is given
by
\begin{IEEEeqnarray*}{l}
\textstyle I( \mathbf s_{}^{(1)}; \mathbf Y_{\!R,L}^{} \,|\, \mathbf
H_{SR}^{}, \mathbf P_{\!S}^{} ) = \log_{2}^{} ( 1 +
\gamma_{\!R,L}^{(1)} ) = \log_{2}^{} \! \bigg(\! 1 +
\frac{\lambda_{SR}^{(1)}}{\lambda_{SR}^{(2)} + N_{0}^{}}
\bigg),\IEEEyesnumber\label{eqn:PerStreamMI1}
\end{IEEEeqnarray*}
and the mutual information for $\textbf{s}_{}^{(2)}$ is given by
\begin{subnumcases}{I( \mathbf s_{}^{(2)}; \mathbf Y_{\!R,L}^{} \,|\,
\mathbf H_{SR}^{}, \mathbf P_{\!S}^{},
\widehat{\textbf{s}}{\,}_{R}^{(1)} ) = \log_{2}^{} ( 1 +
\gamma_{\!R,L}^{(2)} ) =\label{eqn:PerStreamMI2}}
\;\;\;\;\;\log_{2}^{} ( 1 + \lambda_{SR}^{(2)}\,/\!N_{0}^{} )
&\!\!\!\!\!\!if $\textbf{s}_{}^{(1)} =
\widehat{\textbf{s}}{\,}_{R}^{(1)}$,\;\;\;\;\label{eqn:PerStreamMI2Correct}\\
\textstyle\log_{2}^{} \! \bigg(\! 1 + \frac{\lambda_{SR}^{(2)}}{4
\epsilon_{\!R}^{} \lambda_{SR}^{(1)} + N_{0}^{}} \bigg)
&\!\!\!\!\!\!otherwise.\label{eqn:PerStreamMI2Incorrct}
\end{subnumcases}

\textbf{PDF Case 1:} This case results when the relay node correctly
decodes source data stream $\textbf{s}_{}^{(1)}$. Let $D =
2_{}^{L/T} - 1$. As per
\eqref{eqn:PerStreamMI1}-\eqref{eqn:PerStreamMI2} the probability of
PDF~Case~1 is given by
\begin{IEEEeqnarray*}{l}
\textstyle\textrm{Pr} \, (\mathcal A_{1}^{}) = \textrm{Pr} \, \big(
I( \mathbf s_{}^{(1)}; \mathbf Y_{\!R,L}^{} \,|\, \mathbf H_{SR}^{},
\mathbf P_{\!S}^{} ) \ge L/T \big) = \textrm{Pr} \bigg(\!
\frac{\lambda_{SR}^{(1)}}{\lambda_{SR}^{(2)} + N_{0}^{}} \ge D
\bigg).
\end{IEEEeqnarray*}
Data stream $\textbf{s}_{}^{(1)}$ is the cooperative stream and is
forwarded by the relay node to the destination node. Combining the
received signals in the listening and cooperative phases, it can be
shown that the end-to-end mutual information for
$\textbf{s}_{}^{(1)}$ is ${I\!\;}_{\textrm{PDF}}^{(1)} (\mathcal H,
\mathbf P_{\!S}^{}, \mathbf P_{\!R}^{}) = \log_{2}^{} \! \Bigg(\! 1
+ \frac{\big( (\lambda_{SD}^{(1)})_{}^{1\!/2} +
(\lambda_{RD}^{})_{}^{1\!/2} \big)_{}^{2}}{\lambda_{SD}^{(2)} +
2N_{0}^{}} \Bigg)$, where $\lambda_{SD}^{(1)} = || \mathbf H_{SD}^{}
\, [\mathbf P_{\!S}^{}]_{(:, 1:n_{S}^{(1)})}^{} ||_{}^{2}$,
$\lambda_{SD}^{(2)} = || \mathbf H_{SD}^{} \, [\mathbf
P_{\!S}^{}]_{(:, n_{S}^{(1)}+1:n_{\!S}^{})}^{} ||_{}^{2}$, and
$\lambda_{RD}^{} = || \mathbf H_{RD}^{} \, \mathbf P_{\!R}^{}
||_{}^{2}$. Thus, the end-to-end outage probability of
$\textbf{s}_{}^{(1)}$ is
\hbox{$P_{\textrm{out}}^{(1)} (\mathbf P_{\!S}^{}, \mathbf
P_{\!R}^{} \,|\, \mathcal A_{1}^{}) = \textrm{Pr} \Bigg(\!
\frac{\big( (\lambda_{SD}^{(1)})_{}^{1\!/2} +
(\lambda_{RD}^{})_{}^{1\!/2} \big)_{}^{2}}{\lambda_{SD}^{(2)} +
2N_{0}^{}} < D \Bigg)$}. On the other hand, data stream
$\textbf{s}_{}^{(2)}$ is the regular payload stream and is not
forwarded by the relay node. Suppose the cooperative stream
$\textbf{s}_{}^{(1)}$ is decoded as
$\widehat{\textbf{s}}{\,}_{D}^{(1)}$ at SER
$\epsilon_{\mathcal{A}_{1}^{}}^{}\!$ and we cancel its interference
from the SD link received signals. If $\textbf{s}_{}^{(1)} =
\widehat{\textbf{s}}{\,}_{D}^{(1)}$, then the end-to-end mutual
information for $\textbf{s}_{}^{(2)}$ is
${I\!\;}_{\textrm{PDF}}^{(2)} (\mathcal H, \mathbf P_{\!S}^{},
\mathbf P_{\!R}^{}) = \log_{2}^{} ( 1 +
\lambda_{SD}^{(2)}\,/\!N_{0}^{} )$; otherwise, the end-to-end mutual
information for $\textbf{s}_{}^{(2)}$ is
${I\!\;}_{\textrm{PDF}}^{(2)} (\mathcal H, \mathbf P_{\!S}^{},
\mathbf P_{\!R}^{}) = \log_{2}^{} \! \bigg(\! 1 +
\frac{\lambda_{SD}^{(2)}}{4 \, \epsilon_{\!\mathcal{A}_{1}^{}}^{} \!
\lambda_{SD}^{(1)} + N_{0}^{}} \bigg)$. Therefore, the end-to-end
outage probability of $\textbf{s}_{}^{(2)}$ is\\
\centerline{$P_{\textrm{out}}^{(2)} (\mathbf P_{\!S}^{}, \mathbf
P_{\!R}^{} |\, \mathcal A_{1}^{}) \!=\! ( 1 \!-\!
P_{\textrm{out}}^{(1)} (\mathbf P_{\!S}^{}, \mathbf P_{\!R}^{} |\,
\mathcal A_{1}^{}) ) \, \textrm{Pr} \, ( \lambda_{SD}^{(2)} \!<\! D
) \!+\! P_{\textrm{out}}^{(1)} (\mathbf P_{\!S}^{}, \mathbf
P_{\!R}^{} |\, \mathcal A_{1}^{}) \, \textrm{Pr} \bigg(\!
\frac{\lambda_{SD}^{(2)}}{4 \, \epsilon_{\!\mathcal{A}_{1}^{}}^{} \!
\lambda_{SD}^{(1)} + N_{0}^{}} \!<\! D \bigg)$.}

\textbf{PDF~Case~2:} This case results when the relay node
incorrectly decodes source data stream $\mathbf s_{}^{(1)}$ but
correctly decodes data stream $\mathbf s_{}^{(2)}$. As per
\eqref{eqn:PerStreamMI1}-\eqref{eqn:PerStreamMI2}, the probability of PDF~Case~2 is 
\begin{IEEEeqnarray*}{ll}
\textstyle\textrm{Pr} \, (\mathcal A_{2}^{}) \;&\textstyle=
\textrm{Pr} \, \big( I( \mathbf s_{}^{(1)}; \mathbf Y_{\!R,L}^{}
\,|\, \mathbf H_{SR}^{}, \mathbf P_{\!S}^{} ) < L/T \big) \,
\textrm{Pr} \, \big( I( \mathbf s_{}^{(2)}; \mathbf Y_{\!R,L}^{}
\,|\, \mathbf H_{SR}^{}, \mathbf P_{\!S}^{}, \textbf{s}_{}^{(1)}
\neq \widehat{\textbf{s}}{\,}_{R}^{(1)} ) \ge L/T \big)\\
&\textstyle= \textrm{Pr} \bigg(\!
\frac{\lambda_{SR}^{(1)}}{\lambda_{SR}^{(2)} + N_{0}^{}} < D \bigg)
\, \textrm{Pr} \bigg(\! \frac{\lambda_{SR}^{(2)}}{4
\epsilon_{\!R}^{} \lambda_{SR}^{(1)} + N_{0}^{}} \ge D \bigg).
\end{IEEEeqnarray*}
We can derive the outage probabilities of $\textbf{s}_{}^{(1)}$ and
$\textbf{s}_{}^{(2)}$ analogous to PDF~Case~1, but instead we treat
$\textbf{s}_{}^{(1)}$ as the regular payload stream and
$\textbf{s}_{}^{(2)}$ as the cooperative stream.

\textbf{PDF~Case~3:} This case results when the relay node
incorrectly decodes both source data streams $\mathbf s_{}^{(1)}$
and $\mathbf s_{}^{(2)}$. As per
\eqref{eqn:PerStreamMI1}-\eqref{eqn:PerStreamMI2}, the probability
of PDF~Case~3 is given by
\begin{IEEEeqnarray*}{ll}
\textstyle\textrm{Pr} \, (\mathcal A_{3}^{}) \;&\textstyle=
\textrm{Pr} \, \big( I( \mathbf s_{}^{(1)}; \mathbf Y_{\!R,L}^{}
\,|\, \mathbf H_{SR}^{}, \mathbf P_{\!S}^{} ) < L/T \big) \,
\textrm{Pr} \, \big( I( \mathbf s_{}^{(2)}; \mathbf Y_{\!R,L}^{}
\,|\, \mathbf H_{SR}^{}, \mathbf P_{\!S}^{}, \textbf{s}_{}^{(1)}
\neq \widehat{\textbf{s}}{\,}_{R}^{(1)} ) < L/T \big)\\
&\textstyle= \textrm{Pr} \bigg(\!
\frac{\lambda_{SR}^{(1)}}{\lambda_{SR}^{(2)} + N_{0}^{}} < D \bigg)
\, \textrm{Pr} \bigg(\! \frac{\lambda_{SR}^{(2)}}{4
\epsilon_{\!R}^{} \lambda_{SR}^{(1)} + N_{0}^{}} < D \bigg).
\end{IEEEeqnarray*}
Since neither $\mathbf s_{}^{(1)}$ nor $\mathbf s_{}^{(2)}$ is
forwarded by the relay node, at the destination node we decode the
data streams only from the SD link received signals. The mutual
information for each data stream can be determined similar to the
derivation of \eqref{eqn:PerStreamMI1}-\eqref{eqn:PerStreamMI2}
focusing instead on the SD link.
\section*{Appendix B: Proof of Theorem~\ref{Thm:relay}}
\label{pf:Thm_relay} For analytical tractability, we assume that
under PDF Case 1 and 2 the SER of the cooperative stream is
reasonably low (e.g. $\epsilon_{\mathcal{A}_{1}^{}}^{},
\epsilon_{\mathcal{A}_{2}^{}}^{} \ll 1$), and the outage
probabilities of the regular payload stream under PDF Case 1 and 2
can be approximated as:
\begin{IEEEeqnarray*}{l}
P_{\textrm{out}}^{(2)} (\mathbf P_{\!S}^{}, \mathbf P_{\!R}^{} \,|\,
\mathcal A_{1}^{}) \!\approx \textrm{Pr} \, (
\lambda_{SD}^{(2)}\,/\!N_{0}^{} \!< D
)\;\;\textrm{and}\;\;P_{\textrm{out}}^{(1)} (\mathbf P_{\!S}^{},
\mathbf P_{\!R}^{} \,|\, \mathcal A_{2}^{}) \!\approx \textrm{Pr} \,
( \lambda_{SD}^{(1)}\,/\!N_{0}^{} \!< D
).\;\;\;\;\;\;\IEEEyesnumber\label{Eqn:CoopStreamApprox}
\end{IEEEeqnarray*}
Thus, the average end-to-end per-stream outage probability
$P_{\textrm{out}}^{}( \mathbf P_{\!S}^{}, \mathbf P_{\!R}^{} )$ is
related to the relay node precoder $\mathbf P_{\!R}^{}$ only w.r.t.
the probabilities
$P_{\textrm{out}}^{(1)} (\mathbf P_{\!S}^{}, \mathbf P_{\!R}^{}
\,|\, \mathcal A_{1}^{}) \!\!< \textrm{Pr} \Big(\! \lambda_{RD}^{}
\!<\! ( \lambda_{SD}^{(2)} \!+\! 2N_{0}^{})D \!-\!
\lambda_{SD}^{(1)} \Big)$ and $P_{\textrm{out}}^{(2)} (\mathbf
P_{\!S}^{}, \mathbf P_{\!R}^{} \,|\, \mathcal A_{2}^{}) \!\!<
\textrm{Pr} \Big(\! \lambda_{RD}^{} \!<\! ( \lambda_{SD}^{(1)} \!+\!
2N_{0}^{})D \!-\! \lambda_{SD}^{(2)} \Big)$. To minimize
$P_{\textrm{out}}^{(1)} (\mathbf P_{\!S}^{}, \mathbf P_{\!R}^{}
\,|\, \mathcal A_{1}^{})$ and $P_{\textrm{out}}^{(2)} (\mathbf
P_{\!S}^{}, \mathbf P_{\!R}^{} \,|\, \mathcal A_{2}^{})$, we design
$\mathbf P_{\!R}^{}$ to minimize the probability density function
(p.d.f.) of $\lambda_{RD}^{}$.

The p.d.f. of $\lambda_{RD}^{}$ is given as follows. By definition,
$\lambda_{RD}^{} = || \mathbf H_{RD}^{} \, \mathbf P_{\!R}^{}
||_{}^{2} = ( \mathbf g_{RD}^{} )_{}^{\!\dag} \mathbf g_{RD}^{}$,
where $\mathbf g_{RD}^{} = \textrm{vec}( \mathbf H_{RD}^{} \,
\mathbf P_{\!R}^{} ) \stackrel{(a)}{=} ( ( \mathbf P_{\!R}^{}
)_{}^{T} ( \mathbf \Lambda_{RD,t}^{} )_{}^{\!1\!\!\:/\!\:2} \otimes
( \mathbf \Lambda_{RD,r}^{} )_{}^{\!1\!\!\:/\!\:2} ) \,
\textrm{vec}( \mathbf G_{\!RD}^{} )$ and (a) follows from
\eqref{Eqn:ChannelDecompose}. Since the entries of $\mathbf
G_{\!RD}^{}$ are i.i.d. complex Gaussian with zero mean and unit
variance, $\mathbf g_{RD}^{}$ is multivariate Gaussian distributed
whose p.d.f. is given by $ p( \mathbf g_{RD}^{} ) = \frac{\exp(-(
\mathbf g_{RD}^{} )_{}^{\dag} \mathbf \Omega_{}^{-1} \mathbf
g_{RD}^{})}{(2 \pi)_{}^{V} \det ( \mathbf \Omega )}$, where
$V=n_{\!D}^{} n_{\!R}^{}$, $\mathbf \Omega = ( \mathbf P_{\!R}^{}
)_{}^{T} \mathbf \Xi_{RD,t}^{} ( \mathbf P_{\!R}^{} )_{}^{\ast}
\otimes \mathbf \Xi_{RD,r}^{}$, $\mathbf \Xi_{RD,t}^{} = ( \mathbf
\Lambda_{RD,t}^{} )_{}^{\!1\!\!\:/\!\:2} ( ( \mathbf
\Lambda_{RD,t}^{} )_{}^{\!1\!\!\:/\!\:2} )_{}^{\dag}$, and $\mathbf
\Xi_{RD,r}^{} = ( \mathbf \Lambda_{RD,r}^{} )_{}^{\!1\!\!\:/\!\:2} (
( \mathbf \Lambda_{RD,r}^{} )_{}^{\!1\!\!\:/\!\:2} )_{}^{\dag}$.
Let $\mathbf \Omega_{}^{-1} = \textbf{U}_{\mathbf \Omega}^{} (
\mathbf \Sigma_{\mathbf \Omega}^{} )_{}^{-1} ( \textbf{U}_{\mathbf
\Omega}^{} )_{}^{\dag}$ denote the eigendecomposition of $\mathbf
\Omega_{}^{-1}$, where $\mathbf \Sigma_{\mathbf \Omega}^{} =
\textrm{diag}( \delta_{\max}^{}, \ldots, \delta_{\min}^{} )$ with
$\delta_{\max}^{}$ and $\delta_{\min}^{}$ denote the maximum and
minimum eigenvalues of $\mathbf \Omega$. Hence, $( \mathbf g_{RD}^{}
)_{}^{\dag} \mathbf \Omega_{}^{-1} \mathbf g_{RD}^{} \geq ( \mathbf
g_{RD}^{} )_{}^{\dag} \textbf{U}_{\mathbf \Omega}^{} \big(
\frac{1}{\delta_{\max}^{}} \textbf{I}_{V}^{} \big) (
\textbf{U}_{\mathbf \Omega}^{} )_{}^{\dag} \mathbf g_{RD}^{} =
\frac{1}{\delta_{\max}^{}}( \mathbf g_{RD}^{} )_{}^{\dag} \mathbf
g_{RD}^{}$ and the p.d.f. of $\lambda_{RD}^{}$ is upper bounded by $
p( \lambda_{RD}^{} ) \leq
\frac{\exp(-\lambda_{RD}^{}/\delta_{\max}^{})}{(2 \pi)_{}^{V}
(\delta_{\min}^{})_{}^{V}}$.

To minimize the p.d.f. of $\lambda_{RD}^{}$ (and thereby minimize
$P_{\textrm{out}}^{}( \mathbf P_{\!S}^{}, \mathbf P_{\!R}^{} )$), we
seek to minimize $\delta_{\max}^{}$ \emph{and} maximize
$\delta_{\min}^{}$; this implies minimizing the \emph{condition
number} of $\mathbf \Omega$, $\chi(\mathbf \Omega) =
\frac{\delta_{\max}^{}}{\delta_{\min}^{}}$. It can be shown
that\footnote{Let $\sigma_{\max}^{} (\textbf{A})$ and
$\sigma_{\min}^{} (\textbf{A})$ denote, respectively, the maximum
and minimum eigenvalues of of $\textbf{A}$. Since $\sigma_{\max}^{}
(\textbf{A} \otimes \textbf{B}) = \sigma_{\max}^{} (\textbf{A})
\sigma_{\max}^{} (\textbf{B})$ and $\sigma_{\min}^{} (\textbf{A}
\otimes \textbf{B}) = \sigma_{\min}^{} (\textbf{A}) \sigma_{\min}^{}
(\textbf{B})$, so $\chi(\textbf{A} \otimes \textbf{B}) =
\chi(\textbf{A})
\chi(\textbf{B})$.}\newcounter{ConditionNumberFootnote}\setcounter{ConditionNumberFootnote}{\value{footnote}}
$\chi(\mathbf \Omega) = \chi(( \mathbf P_{\!R}^{} )_{}^{T} \mathbf
\Xi_{RD,t}^{} ( \mathbf P_{\!R}^{} )_{}^{\ast} \otimes \mathbf
\Xi_{RD,r}^{}) = \chi(( \mathbf P_{\!R}^{} )_{}^{T} \mathbf
\Xi_{RD,t}^{} ( \mathbf P_{\!R}^{} )_{}^{\ast}) \chi(\mathbf
\Xi_{RD,r}^{})$. Therefore, we recast subproblem
\eqref{Eqn:OptProbSubP_R} as
\begin{IEEEeqnarray*}{l}
\min_{\mathbf P_{\!R}^{}} \chi(( \mathbf P_{\!R}^{} )_{}^{T} \mathbf
\Xi_{RD,t}^{} ( \mathbf P_{\!R}^{} )_{}^{\ast}),
\;\;\textrm{s.t.}\;\; || \mathbf P_{\!R}^{} ||_{}^{2} \leq
\alpha_{R}^{}.\IEEEyesnumber\label{Eqn:OptProbSubP_R_Recast}
\end{IEEEeqnarray*}
Let $\mathbf \Xi_{RD,t}^{} = \mathbf U_{\!RD,t}^{} \mathbf
\Sigma_{\!RD,t}^{} ( \mathbf U_{\!RD,t}^{} )_{}^{\dag} $ denote the
eigendecomposition of $\mathbf \Xi_{RD,t}^{}$, and the solution to
\eqref{Eqn:OptProbSubP_R_Recast} is given by $\mathbf
P_{\!R}^{\star} = \sqrt{\frac{\alpha_{R}^{}}{\textrm{Tr} \,((
\mathbf \Sigma_{\!RD,t}^{} )_{}^{-1})}} \, ( \mathbf U_{\!RD,t}^{}
)_{}^{\ast} \, ( \mathbf \Sigma_{\!RD,t}^{} )_{}^{-1/2}$. The
physical meaning of this precoder design is to \emph{equalize} the
RD link transmit-side correlation matrix.

\section*{Appendix C: Proof of
Theorem~\ref{Thm:source}} \label{pf:Thm_source} The probability of
PDF~Case~3 can be upper bounded as
\begin{IEEEeqnarray*}{ll}\label{Eqn:Case3UpperBound}
\textrm{Pr} \, (\mathcal A_{3}^{}) \;&\textstyle= \textrm{Pr}
\bigg(\! \frac{\lambda_{SR}^{(1)}}{\lambda_{SR}^{(2)} + N_{0}^{}} <
2_{}^{L/T} - 1 \bigg) \, \textrm{Pr} \bigg(\!
\frac{\lambda_{SR}^{(2)}}{4 \epsilon_{\!R}^{} \lambda_{SR}^{(1)} +
N_{0}^{}} < 2_{}^{L/T} - 1 \bigg)\\
&\textstyle <C(\epsilon_{R}^{}, D) \textrm{Pr} \Big(\!
\frac{\lambda_{SR}^{(2)}}{4 \epsilon_{\!R}^{} D} - \frac{N_{0}^{}}{4
\epsilon_{\!R}^{}} < \lambda_{SR}^{(1)} <
\lambda_{SR}^{(2)} D + D \Big)\IEEEyessubnumber\label{Eqn:Case3UpperBound1}\\
&\textstyle= C(\epsilon_{R}^{}, D) \int_{k=0}^{\infty} \textrm{Pr}
\Big(\! \frac{k}{4 \epsilon_{\!R}^{} D} - \frac{N_{0}^{}}{4
\epsilon_{\!R}^{}} < \lambda_{SR}^{(1)} < k D + D \Big) \textrm{Pr}
( \lambda_{SR}^{(2)} = k )
dk,\IEEEyessubnumber\label{Eqn:Case3UpperBound2}
\end{IEEEeqnarray*}
where $D = 2_{}^{L/T} - 1$ is the data rate, and $C(\epsilon_{R}^{},
D)$ is a constant that is a function of the SER $\epsilon_{R}^{}$
and $D$. To minimize \eqref{Eqn:Case3UpperBound}, we seek to design
the source precoder $\mathbf P_{\!S}^{}$ to minimize the inner
probability expression for given $k$:
\begin{IEEEeqnarray*}{l}
\min_{\mathbf P_{\!S}^{}} \textstyle \textrm{Pr} \Big(\! \frac{k}{4
\epsilon_{\!R}^{} D} - \frac{N_{0}^{}}{4 \epsilon_{\!R}^{}} <
\lambda_{SR}^{(1)} < k D + D \Big) \textrm{Pr} ( \lambda_{SR}^{(2)}
= k ), \;\;\textrm{s.t.}\;\; || \mathbf P_{\!S}^{} ||_{}^{2} \leq
\alpha_{S}^{}.\IEEEyesnumber\label{Eqn:OptProbSubP_S_Recast}
\end{IEEEeqnarray*}

We solve \eqref{Eqn:OptProbSubP_S_Recast} by first deriving the
precoder structure that minimizes the joint p.d.f. of
$\lambda_{SR}^{(1)}$ and $\lambda_{SR}^{(2)}$, and the p.d.f. is
given as follows. By definition, $\lambda_{SR}^{(1)} = || \mathbf
H_{SR}^{} \, [\mathbf P_{\!S}^{}]_{(:, 1:n_{S}^{(1)})}^{} ||_{}^{2}$
and $\lambda_{SR}^{(2)} = || \mathbf H_{SR}^{} \, [\mathbf
P_{\!S}^{}]_{(:, n_{S}^{(1)}+1:n_{\!S}^{})}^{} ||_{}^{2}$; let
$\lambda_{SR}^{} = \lambda_{SR}^{(1)} + \lambda_{SR}^{(2)} = ||
\mathbf H_{SR}^{} \, \mathbf P_{\!S}^{} ||_{}^{2} = ( \mathbf
g_{SR}^{} )_{}^{\!\dag} \mathbf g_{SR}^{}$, where $\mathbf g_{RD}^{}
= \textrm{vec}( \mathbf H_{RD}^{} \, \mathbf P_{\!R}^{} )$. It can
be shown, in analogy to the proof of Theorem~\ref{Thm:relay}, that
the p.d.f. of $\mathbf g_{SR}^{}$ is $ p( \mathbf g_{SR}^{} ) =
\frac{\exp(-( \mathbf g_{SR}^{} )_{}^{\dag} \mathbf \Phi_{}^{-1}
\mathbf g_{SR}^{})}{(2 \pi)_{}^{W} \det ( \mathbf \Phi )}$, where
$W=n_{\!R}^{} n_{\!S}^{}$, $\mathbf \Phi = ( \mathbf P_{\!S}^{}
)_{}^{T} \mathbf \Xi_{SR,t}^{} ( \mathbf P_{\!S}^{} )_{}^{\ast}
\otimes \mathbf \Xi_{SR,r}^{}$, $\mathbf \Xi_{SR,t}^{} = ( \mathbf
\Lambda_{SR,t}^{} )_{}^{\!1\!\!\:/\!\:2} ( ( \mathbf
\Lambda_{SR,t}^{} )_{}^{\!1\!\!\:/\!\:2} )_{}^{\dag}$, and $\mathbf
\Xi_{SR,r}^{} = ( \mathbf \Lambda_{SR,r}^{} )_{}^{\!1\!\!\:/\!\:2} (
( \mathbf \Lambda_{SR,r}^{} )_{}^{\!1\!\!\:/\!\:2} )_{}^{\dag}$. Let
$\mathbf \Xi_{SR,t}^{} = \mathbf U_{\!SR,t}^{} \mathbf
\Sigma_{\!SR,t}^{} ( \mathbf U_{\!SR,t}^{} )_{}^{\dag} $ denote the
eigendecomposition of $\mathbf \Xi_{SR,t}^{}$. Without loss of
generality, let the source node precoder be given by
\begin{IEEEeqnarray}{l}
\mathbf P_{\!S}^{} = \sqrt{\alpha_{S}^{}} \, ( \mathbf U_{\!SR,t}^{}
)_{}^{\ast} \, \mathbf
\Sigma_{S}^{},\IEEEyesnumber\label{Eqn:SrcPrecoderStructure}
\end{IEEEeqnarray}
where $\mathbf \Sigma_{S}^{}$ is a diagonal matrix. Thus, $\mathbf
\Phi = \alpha_{S}^{} \mathbf \Sigma_{S}^{} \mathbf
\Sigma_{\!SR,t}^{} \mathbf \Sigma_{S}^{} \otimes \mathbf
\Xi_{SR,r}^{} = \Bigg[\!\!
\begin{array}{cc}\mathbf
\Phi_{}^{(1)}\!\!\!\!& \textbf{0}\\
\textbf{0}\!\!\!\!& \mathbf \Phi_{}^{(2)}\end{array}\!\!\Bigg]$ with
$\mathbf \Phi_{}^{(1)} = \alpha_{S}^{} [ \mathbf \Sigma_{S}^{}
\mathbf \Sigma_{\!SR,t}^{} \mathbf \Sigma_{S}^{} ]_{(1:n_{S}^{(1)},
1:n_{S}^{(1)})}^{} \otimes \mathbf \Xi_{SR,r}^{}$ and $\mathbf
\Phi_{}^{(2)} = \alpha_{S}^{} [ \mathbf \Sigma_{S}^{} \mathbf
\Sigma_{\!SR,t}^{} \mathbf \Sigma_{S}^{}
]_{(n_{S}^{(1)}+1:n_{\!S}^{}, n_{S}^{(1)}+1:n_{\!S}^{})}^{} \otimes
\mathbf \Xi_{SR,r}^{}$, and the p.d.f. of $\mathbf g_{SR}^{}$ can be
expressed as
\begin{IEEEeqnarray*}{l}
\textstyle p( \mathbf g_{SR}^{} ) = \frac{\exp(-( [\mathbf
g_{SR}^{}]_{(1:W_{}^{(1)})}^{} )_{}^{\dag} ( \mathbf \Phi_{}^{(1)}
)_{}^{-1} [\mathbf g_{SR}^{}]_{(1:W_{}^{(1)})}^{})}{(2
\pi)_{}^{W_{}^{(1)}} \det ( \mathbf \Phi_{}^{(1)} )} \frac{\exp(-(
[\mathbf g_{SR}^{}]_{(W_{}^{(1)}+1:W)}^{} )_{}^{\dag} ( \mathbf
\Phi_{}^{(2)} )_{}^{-1} [\mathbf
g_{SR}^{}]_{(W_{}^{(1)}+1:W)}^{})}{(2 \pi)_{}^{W_{}^{(2)}}\det (
\mathbf \Phi_{}^{(2)} )}
\end{IEEEeqnarray*}
for $W_{}^{(1)} = n_{\!R}^{}n_{\!S}^{(1)}$ and $W_{}^{(2)} =
n_{\!R}^{}n_{\!S}^{(2)}$. Let $( \mathbf \Phi_{}^{(i)} )_{}^{-1} =
\textbf{U}_{\mathbf \Phi_{}^{(1)}}^{} ( \mathbf \Sigma_{\mathbf
\Phi_{}^{(i)}}^{} )_{}^{-1} ( \textbf{U}_{\mathbf \Phi_{}^{(i)}}^{}
)_{}^{\dag}$ denote the eigendecomposition of $( \mathbf
\Phi_{}^{(i)} )_{}^{-1}$, where $\mathbf \Sigma_{\mathbf
\Phi_{}^{(i)}}^{} = \textrm{diag}( \delta_{\max}^{(i)}, \ldots,
\delta_{\min}^{(i)} )$ with $\delta_{\max}^{(i)}$ and
$\delta_{\min}^{(i)}$ denote, respectively, the the maximum and
minimum eigenvalues of $\mathbf \Phi_{}^{(i)}$. Therefore, the joint
p.d.f. of $\lambda_{SR}^{(1)}$ and $\lambda_{SR}^{(2)}$ is upper
bounded by $ p( \lambda_{SR}^{(1)}, \lambda_{SR}^{(2)} ) = p(
\lambda_{SR}^{(1)} ) p( \lambda_{SR}^{(2)} )$, where $p(
\lambda_{SR}^{(1)} ) \leq
\frac{\exp(-\lambda_{SR}^{(1)}/\delta_{\max}^{(1)})}{(2
\pi)_{}^{W_{}^{(1)}} (\delta_{\min}^{(1)})_{}^{W_{}^{(1)}}}$ and $p(
\lambda_{SR}^{(2)} ) \leq \frac{\exp(
-\lambda_{SR}^{(2)}/\delta_{\max}^{(2)})}{(2 \pi)_{}^{W_{}^{(2)}}
(\delta_{\min}^{(2)})_{}^{W_{}^{(2)}}}$.

Since the joint p.d.f. of $\lambda_{SR}^{(1)}$ and
$\lambda_{SR}^{(2)}$ is \emph{separable}, we \emph{decompose} and
the precoder design problem \eqref{Eqn:OptProbSubP_S_Recast} as
follows. Let $\rho_{S}^{(1)}, \rho_{S}^{(2)} \ge 0$, $\rho_{S}^{(1)}
+ \rho_{S}^{(2)} = 1$, be power allocation variables (which we
address subsequently), and we can recast
\eqref{Eqn:OptProbSubP_S_Recast} as
\begin{IEEEeqnarray*}{l}\label{Eqn:OptProbSubP_S_Recast2}
\min_{\mathbf P_{\!S}^{}} \textstyle \textrm{Pr} \Big(\! \frac{k}{4
\epsilon_{\!R}^{} D} - \frac{N_{0}^{}}{4 \epsilon_{\!R}^{}} <
\lambda_{SR}^{(1)} < k D + D \Big), \;\;\textrm{s.t.}\;\; ||
[\mathbf P_{\!S}^{}]_{(:, 1:n_{S}^{(1)})}^{} ||_{}^{2} \leq
\rho_{S}^{(1)}
\alpha_{S}^{},\IEEEyessubnumber\label{Eqn:OptProbSubP_S_Recast2a}\\
\min_{\mathbf P_{\!S}^{}} \textstyle \textrm{Pr} (
\lambda_{SR}^{(2)} = k ), \;\;\textrm{s.t.}\;\; || [\mathbf
P_{\!S}^{}]_{(:, n_{S}^{(1)}+1:n_{\!S}^{})}^{} ||_{}^{2} \leq
\rho_{S}^{(2)}
\alpha_{S}^{},\IEEEyessubnumber\label{Eqn:OptProbSubP_S_Recast2b}
\end{IEEEeqnarray*}
To solve \eqref{Eqn:OptProbSubP_S_Recast2a} we seek to minimize the
condition number of $\mathbf \Phi_{}^{(1)}$, and similarly to solve
\eqref{Eqn:OptProbSubP_S_Recast2b} we seek to minimize the condition
number of $\mathbf \Phi_{}^{(2)}$. As per
\eqref{Eqn:SrcPrecoderStructure}, the solution to
\eqref{Eqn:OptProbSubP_S_Recast2} is given by
\begin{IEEEeqnarray*}{l}
\mathbf P_{\!S}^{\star} = \sqrt{\alpha_{S}^{}} \, ( \mathbf
U_{\!SR,t}^{} )_{}^{\ast} \! \left[\!\!
\begin{array}{cc}\sqrt{\frac{\rho_{S}^{(1)}}{\textrm{Tr} \,((
\mathbf \Sigma_{\!SR,t}^{(1)} )_{}^{-1})}} ( \mathbf
\Sigma_{\!SR,t}^{(1)})_{}^{-1/2}\!\!\!\!\!\!& \textbf{0}\\
\textbf{0}\!\!\!\!\!\!& \sqrt{\frac{\rho_{S}^{(2)}}{\textrm{Tr} \,((
\mathbf \Sigma_{\!SR,t}^{(2)} )_{}^{-1})}} ( \mathbf
\Sigma_{\!SR,t}^{(2)})_{}^{-1/2}\end{array}\!\!\right],\IEEEyesnumber\label{Eqn:P_SStructureProof}
\end{IEEEeqnarray*}
where $\mathbf \Sigma_{\!SR,t}^{(1)}$ = $[\mathbf
\Sigma_{\!SR,t}^{}]_{(1:n_{S}^{(1)},1:n_{S}^{(1)})}^{}$ and $\mathbf
\Sigma_{\!SR,t}^{(2)}$ = $[\mathbf
\Sigma_{\!SR,t}^{}]_{(n_{S}^{(1)}+1:n_{\!S}^{},
n_{S}^{(1)}+1:n_{\!S}^{})}^{}$.

Given the precoder structure in \eqref{Eqn:P_SStructureProof}, we
recast \eqref{Eqn:OptProbSubP_S_Recast} to solve for the power
allocation variables $\rho_{S}^{(1)}$ and $\rho_{S}^{(2)}$. The
probabilities in \eqref{Eqn:OptProbSubP_S_Recast2} are given by
\begin{IEEEeqnarray*}{l}\label{Eqn:OptProbSubP_SFinalProbs}
\textstyle \textrm{Pr} \Big(\! \frac{k}{4 \epsilon_{\!R}^{} D} \!-\!
\frac{N_{0}^{}}{4 \epsilon_{\!R}^{}} \!<\! \lambda_{SR}^{(1)} \!<\!
k D \!+\! D \Big) \leq \frac{\delta_{\max}^{(1)}}{(2
\pi)_{}^{W_{}^{(1)}} (\delta_{\min}^{(1)})_{}^{W_{}^{(1)}}} \left(\!
\exp\!\left(\!-\frac{\frac{k}{4 \epsilon_{\!R}^{} D} -
\frac{N_{0}^{}}{4 \epsilon_{\!R}^{}}}{\delta_{\max}^{(1)}}\right) -
\exp\!\left(\!-\frac{k D + D}{\delta_{\max}^{(1)}}\right)
\!\right),\;\;\;\;\;\;\IEEEyessubnumber\label{Eqn:OptProbSubP_SFinalProbsA}\\
\textstyle \textrm{Pr} ( \lambda_{SR}^{(2)} = k ) \leq \frac{\exp(
-k/\delta_{\max}^{(2)})}{(2 \pi)_{}^{W_{}^{(2)}}
(\delta_{\min}^{(2)})_{}^{W_{}^{(2)}}},\IEEEyessubnumber\label{Eqn:OptProbSubP_SFinalProbsB}
\end{IEEEeqnarray*}
where\footnotemark[\value{ConditionNumberFootnote}]
$\delta_{\max}^{(1)} = \frac{\rho_{S}^{(1)}
\alpha_{S}^{}\sigma_{\max}^{} (\mathbf \Xi_{SR,r}^{})}{\textrm{Tr}
\,(( \mathbf \Sigma_{\!SR,t}^{(1)} )_{}^{-1})}$,
$\delta_{\min}^{(1)} = \frac{\rho_{S}^{(1)}
\alpha_{S}^{}\sigma_{\min}^{} (\mathbf \Xi_{SR,r}^{})}{\textrm{Tr}
\,(( \mathbf \Sigma_{\!SR,t}^{(1)} )_{}^{-1})}$,
$\delta_{\max}^{(2)} = \frac{\rho_{S}^{(2)}
\alpha_{S}^{}\sigma_{\max}^{} (\mathbf \Xi_{SR,r}^{})}{\textrm{Tr}
\,(( \mathbf \Sigma_{\!SR,t}^{(2)} )_{}^{-1})}$, and
$\delta_{\min}^{(2)} = \frac{\rho_{S}^{(2)}
\alpha_{S}^{}\sigma_{\min}^{} (\mathbf \Xi_{SR,r}^{})}{\textrm{Tr}
\,(( \mathbf \Sigma_{\!SR,t}^{(2)} )_{}^{-1})}$. Substituting
\eqref{Eqn:OptProbSubP_SFinalProbs} into
\eqref{Eqn:OptProbSubP_S_Recast}, and let
$\widetilde{\delta}_{}^{(1)} =
\frac{\delta_{\max}^{(1)}}{\rho_{S}^{(1)}}$ and
$\widetilde{\delta}_{}^{(2)} =
\frac{\delta_{\max}^{(2)}}{\rho_{S}^{(2)}}$, we have
\begin{IEEEeqnarray*}{l}
\min_{\rho_{S}^{(1)}, \rho_{S}^{(2)} \ge 0} \textstyle
\frac{\exp\left(-\frac{k}{\rho_{S}^{(2)}\widetilde{\delta}_{}^{(2)}}\right)}{(\rho_{S}^{(2)})_{}^{W_{}^{(2)}}(\rho_{S}^{(2)})_{}^{W_{}^{(1)}-1}}
\left(\! \exp\!\left(\!-\frac{\frac{k}{4 \epsilon_{\!R}^{} D} -
\frac{N_{0}^{}}{4
\epsilon_{\!R}^{}}}{\rho_{S}^{(1)}\widetilde{\delta}_{}^{(1)}}\right)
- \exp\!\left(\!-\frac{k D +
D}{\rho_{S}^{(1)}\widetilde{\delta}_{}^{(1)}}\right)
\!\right)\IEEEyessubnumber\label{Eqn:OptProbSubP_S_Power1}\\
\;\;\;\;\textrm{s.t.}\;\;\;\;\; \rho_{S}^{(1)} \!+\! \rho_{S}^{(2)}
\!\leq\! 1.\;\;\IEEEyessubnumber\label{Eqn:OptProbSubP_S_Power2}
\end{IEEEeqnarray*}
Finally, we can determine $\rho_{S}^{(1)}$ and $\rho_{S}^{(2)}$ as
follows. Substitute $\rho_{S}^{(2)} = 1 - \rho_{S}^{(1)}$ into
\eqref{Eqn:OptProbSubP_S_Power1} and take the first order derivative
w.r.t. $\rho_{S}^{(1)}$ whose roots give the optimal value of
$\rho_{S}^{(1)}$ and they can be found using, for example, bisection
search or Newton's algorithm \cite{Boyd03}. After that, it is
straightforward to determine $\rho_{S}^{(2)}$.

\bibliographystyle{IEEEtran}
\bibliography{IEEEabrv,mybibfile}

\begin{thebibliography}{10}
\providecommand{\url}[1]{#1}
\csname url@samestyle\endcsname
\providecommand{\newblock}{\relax}
\providecommand{\bibinfo}[2]{#2}
\providecommand{\BIBentrySTDinterwordspacing}{\spaceskip=0pt\relax}
\providecommand{\BIBentryALTinterwordstretchfactor}{4}
\providecommand{\BIBentryALTinterwordspacing}{\spaceskip=\fontdimen2\font plus
\BIBentryALTinterwordstretchfactor\fontdimen3\font minus
  \fontdimen4\font\relax}
\providecommand{\BIBforeignlanguage}[2]{{%
\expandafter\ifx\csname l@#1\endcsname\relax
\typeout{** WARNING: IEEEtran.bst: No hyphenation pattern has been}%
\typeout{** loaded for the language `#1'. Using the pattern for}%
\typeout{** the default language instead.}%
\else
\language=\csname l@#1\endcsname
\fi
#2}}
\providecommand{\BIBdecl}{\relax}
\BIBdecl

\bibitem{Laneman04}
J.~N. Laneman, D.~N.~C. Tse, and G.~W. Wornell, ``Cooperative diversity in
  wireless networks: Efficient protocols and outage behavior,'' \emph{{IEEE}
  Trans. Inf. Theory}, vol.~50, pp. 3062--3080, Dec. 2004.

\bibitem{Avestimehr07}
A.~S. Avestimehr and D.~N.~C. Tse, ``Outage capacity of the fading relay
  channel in the low {SNR} regime,'' \emph{{IEEE} Trans. Inf. Theory}, vol.~53,
  pp. 1401--1415, Apr. 2007.

\bibitem{Azarian05}
K.~Azarian, H.~E. Gamal, and P.~Schniter, ``On the achievable diversity
  multiplexing tradeoff in half-duplex cooperative channels,'' \emph{{IEEE}
  Trans. Inf. Theory}, vol.~51, pp. 4152--4172, Dec. 2005.

\bibitem{Yuksel04}
M.~Yuksel and E.~Erkip, ``Broadcast strategies for the fading relay channel,''
  in \emph{Proc. {IEEE} {MILCOM}'04}, 2004.

\bibitem{Ding08}
Y.~Ding, J.-K. Zhang, and K.~Wong, ``Optimal precoder for amplify-and-forward
  half-duplex relay system,'' \emph{{IEEE} Trans. Wireless Commun.}, vol.~7,
  pp. 2890--2895, Aug. 2008.

\bibitem{Jnl:RelayPrecoderOnly}
X.~Tang and Y.~Hua, ``Optimal design of non-regenerative {MIMO} wireless
  relays,'' \emph{{IEEE} Trans. Wireless Commun.}, vol.~6, pp. 1398--1407, Apr.
  2007.

\bibitem{Cnf:RelayPrecoderOnly}
O.~Munoz, J.~Vidal, and A.~Agustin, ``Non-regenerative {MIMO} relaying with
  channel state information,'' in \emph{Proc. IEEE ICASSP'05}, Mar. 2005.

\bibitem{ImperfectCSIRelayPrecoder:HKU}
C.~Xing, S.~Ma, and Y.-C. Wu, ``Robust joint design of linear relay precoder
  and destination equalizer for dual-hop amplify-and-forward {MIMO} relay
  systems,'' \emph{{IEEE} Trans. Signal Process.}, vol.~58, pp. 2273--2283,
  Apr. 2010.

\bibitem{Khoshnevis08}
B.~Khoshnevis, W.~Yu, and R.~Adve, ``Grassmannian beamforming for {MIMO}
  amplify-and-forward relaying,'' \emph{{IEEE} J. Sel. Areas Commun.}, vol.~26,
  pp. 1397--1407, Oct. 2008.

\bibitem{Lokesh08}
S.~S. Lokesh, A.~Kumar, and M.~Agrawal, ``Structure of an optimum linear
  precoder and its application to {ML} equalizer,'' \emph{{IEEE} Trans. Signal
  Process.}, vol.~56, pp. 3690--3701, Aug. 2008.

\bibitem{Sezgin07}
A.~Sezgin, A.~Paulraj, and M.~Vu, ``Impact of correlation on linear precoding
  in {QSTBC} coded systems with linear {MSE} detection,'' in \emph{Proc. {IEEE}
  {GLOBECOM}'07}, Nov. 2007.

\bibitem{Meng07}
C.~Meng and J.~Tuqan, ``Precoded {STBC-VBLAST} for {MIMO} wireless
  communication systems,'' in \emph{Proc. {IEEE} {ICASSP}'07}, 2007.

\bibitem{Alamouti98}
S.~Alamouti, ``A simple transmit diversity technique for wireless
  communications,'' \emph{{IEEE} J. Sel. Areas Commun.}, vol.~16, pp.
  1451--1458, Oct. 1998.

\bibitem{OSTBC}
V.~Tarokh, H.~Jafarkhani, and A.~R. Calderbank, ``Space-time block codes from
  orthogonal designs,'' \emph{{IEEE} Trans. Inf. Theory}, vol.~45, pp.
  744--765, Jul. 1999.

\bibitem{QOSTBC}
H.~Jafarkhani, ``A quasi orthogonal space-time block code,'' \emph{{IEEE}
  Trans. Commun.}, vol.~49, pp. 1--4, Jan. 2001.

\bibitem{Jnl:RobustQosP2PMimo:Palomar}
A.~Pascual-Iserte, D.~P. Palomar, A.~I. Prez-Neira, and M.~A. Lagunas, ``A
  robust maximin approach for {MIMO} communications with partial channel state
  information based on convex optimization,'' \emph{{IEEE} Trans. Signal
  Process.}, vol.~54, pp. 346--360, Jan. 2006.

\bibitem{Std:16m}
\emph{{Draft Amendment to IEEE Standard for Local and Metropolitan Area
  Networks, Part 16: Air Interface for Fixed and Mobile Broadband Wireless
  Access Systems}}, IEEE Std. P802.16m/D10, 2010.

\bibitem{Weichselberger06}
W.~Weichselberger, M.~Herlin, H.~Ozcelik, and E.~Bonek, ``A stochastic {MIMO}
  channel model with joint correlation of both link ends,'' \emph{{IEEE} Trans.
  Wireless Commun.}, vol.~5, pp. 90--100, Jan. 2006.

\bibitem{Foschini99}
G.~J. Foschini, G.~D. Golden, R.~A. Valenzuela, and P.~W. Wolniansky,
  ``Simplified processing for high spectral efficiency wireless communication
  employing multi-element arrays,'' \emph{{IEEE} Trans. Commun.}, vol.~17, pp.
  1841--1852, Nov. 1999.

\bibitem{Jnl:SumPowerConstraint1}
S.~Senthuran, A.~Anpalagan, and O.~Das, ``Cooperative subcarrier and power
  allocation for a two-hop decode-and-forward {OFCDM} based relay network,''
  \emph{{IEEE} Trans. Wireless Commun.}, vol.~8, pp. 4797--4805, Sep. 2009.

\bibitem{Jnl:SumPowerConstraint2}
M.~Chen, S.~Serbetli, and A.~Yener, ``Distributed power allocation strategies
  for parallel relay networks,'' \emph{{IEEE} Trans. Wireless Commun.}, vol.~7,
  pp. 552--561, Feb. 2008.

\bibitem{Cnf:SumPowerConstraint}
N.~Zhou, X.~Zhu, Y.~Huang, and H.~Lin, ``Adaptive resource allocation for
  multi-destination relay systems based on {OFDM} modulation,'' in \emph{Proc.
  IEEE ICC'09}, Jun. 2009.

\bibitem{Palomar03}
D.~P. Palomar, J.~M. Cioffi, and M.~A. Lagunas, ``Joint {Tx-Rx} beamforming
  design for multicarrier {MIMO} channels: A unified framework for convex
  optimization,'' \emph{{IEEE} Trans. Signal Process.}, vol.~51, pp.
  2381--2401, Sep. 2003.

\bibitem{Jnl:PrecoderStatCsit1}
H.~Sampath and A.~Paulraj, ``Linear precoding for space-time coded systems with
  known fading correlations,'' \emph{{IEEE} Commun. Lett.}, vol.~6, pp.
  239--241, Jun. 2002.

\bibitem{Jnl:PrecoderStatCsit2}
H.~R. Bahrami and T.~Le-Ngoc, ``Precoder design based on correlation matrices
  for mimo systems,'' \emph{{IEEE} Trans. Wireless Commun.}, vol.~5, pp.
  3579--3587, Dec. 2006.

\bibitem{Boyd03}
S.~Boyd and L.~Vandenberghe, \emph{Convex Optimization}.\hskip 1em plus 0.5em
  minus 0.4em\relax Cambridge University Press, 2003.

\bibitem{Std:4G-EMD:16m}
{IEEE} 802.16m evaluation methodology document. {IEEE} 802.16m-08/004r4.

\bibitem{Non-OrthogonalRelay}
R.~U. Nabar, H.~B$\ddot{\textrm{o}}$lcskei, and F.~W.
  Kneub$\ddot{\textrm{u}}$hler, ``Fading relay channels: Performance limits and
  space-time signal design,'' \emph{{IEEE} J. Sel. Areas Commun.}, vol.~22, pp.
  1099--1109, Aug. 2004.

\bibitem{Ganesan01}
G.~Ganesan and P.~Stoica, ``Space-time block codes: A maximum {SNR} approach,''
  \emph{{IEEE} Trans. Inf. Theory}, vol.~47, pp. 1650--1656, May 2001.

\end{thebibliography}

\newpage

\end{document}